\documentclass[fleqn,11pt,twoside]{article}

\usepackage{amsthm,amsthm,amssymb, color, xcolor,epsfig, graphics, subfigure}

\usepackage{amsmath, graphicx, latexsym, lscape }

\usepackage{cite}

\makeatletter
\newcommand{\copyrightnote}[2]{{\renewcommand{\thefootnote}{}
 \footnotetext{\small\it
\begin{flushleft}
 \copyright \ #1   #2
\end{flushleft}}}}

\newcommand{\Name}[1]{\begin{flushleft}
                       \LARGE \bf #1
                       \end{flushleft}\vspace{-3mm}}

\newcommand{\Author}[1]{\begin{flushleft}
                       \it #1 \end{flushleft}}

\newcommand{\Address}[1]{\begin{flushleft}
                       \it #1 \end{flushleft}}

\newcommand{\Date}[1]{\begin{flushleft}
                      \small  \it #1 \end{flushleft}}

\newcommand{\evenhead}{Author \ name}
\newcommand{\oddhead}{Article \ name}

\renewcommand{\@evenhead}{
\hspace*{-3pt}\raisebox{-15pt}[\headheight][0pt]{\vbox{\hbox to \textwidth
{\thepage \hfil \evenhead}\vskip4pt \hrule}}}
\renewcommand{\@oddhead}{
\hspace*{-3pt}\raisebox{-15pt}[\headheight][0pt]{\vbox{\hbox to \textwidth
{\oddhead \hfil \thepage}\vskip4pt\hrule}}}
\renewcommand{\@evenfoot}{}
\renewcommand{\@oddfoot}{}

\long\def\@makecaption#1#2{%
  \vskip\abovecaptionskip
  \sbox\@tempboxa{\small \textbf{#1.}\ \ #2}%
  \ifdim \wd\@tempboxa >\hsize
    {\small \textbf{#1.}\ \ #2}\par
  \else
    \global \@minipagefalse
    \hb@xt@\hsize{\hfil\box\@tempboxa\hfil}%
  \fi
  \vskip\belowcaptionskip}

\newcommand{\JNMPnumberwithin}[3][\arabic]{%
  \@ifundefined{c@#2}{\@nocounterr{#2}}{%
    \@ifundefined{c@#3}{\@nocnterr{#3}}{%
      \@addtoreset{#2}{#3}%
      \@xp\xdef\csname the#2\endcsname{%
        \@xp\@nx\csname the#3\endcsname .\@nx#1{#2}}}}%
}

\newcommand{\resetfootnoterule} {
  \renewcommand\footnoterule{%
  \kern-3\p@
  \hrule\@width.4\columnwidth
  \kern2.6\p@}
}

\renewcommand{\footnoterule}{}

\makeatother

\theoremstyle{definition}

\def\a{\alpha}
\def\b{\beta}
\def\ga{\gamma}

\def\vphi{\varphi}

\def\s{\sigma}

\def\vphi{\varphi}

\def\De{\Delta}

\def\pa{\partial}

\def\o+{\oplus}

\def\<{\langle}
\def\>{\rangle}

\def\Lap{\Delta}

\def\xt{{\wt{x}}}

\def\({\left(}
\def\){\right)}
\def\[{\left[}
\def\]{\right]}
\def\=#1{\bar #1}
\def\~#1{\widetilde #1}
\def\wt#1{\widetilde #1}
\def\.#1{\dot #1}
\def\^#1{\widehat #1}

\def\"#1{\ddot #1}

\def\eeq{\end{equation}}
\def\beq{\begin{equation}}

\def\beql#1{\begin{equation} \label{#1}}

\def\eqref#1{(\ref{#1})}

\def\EOR{ \hfill $\odot$ \medskip}

\def\symmref{AVL,CGbook,KrV,Olver1,Olver2,Stephani}
\def\sderef{Arnold,Evans,Fre,Ikeda,Kampen,Oksendal,Stroock}
\def\kozref{Koz1,Koz2,Koz3,Koz2018,Koz18a,Koz18b}

\begin{document}

\renewcommand{\evenhead}{ {\LARGE\textcolor{blue!10!black!40!green}{{\sf \ \ \ ]ocnmp[}}}\strut\hfill G. Gaeta \& M.A. Rodr\'iguez}
\renewcommand{\oddhead}{ {\LARGE\textcolor{blue!10!black!40!green}{{\sf ]ocnmp[}}}\ \ \ \ \ Symmetry classification of scalar Ito equations}

\thispagestyle{empty}
\newcommand{\FistPageHead}[3]{
\begin{flushleft}
\raisebox{8mm}[0pt][0pt]
{\footnotesize \sf
\parbox{150mm}{{Open Communications in Nonlinear Mathematical Physics}\ \  \ {\LARGE\textcolor{blue!10!black!40!green}{]ocnmp[}}
\ \ Vol.2 (2022) pp
#2\hfill {\sc #3}}}\vspace{-13mm}
\end{flushleft}}

\FistPageHead{1}{\pageref{firstpage}--\pageref{lastpage}}{ \ \ Article}

\strut\hfill

\strut\hfill

\copyrightnote{The author(s). Distributed under a Creative Commons Attribution 4.0 International License}

\Name{Symmetry classification of scalar autonomous Ito stochastic differential equations with simple noise}

\Author{G. Gaeta$^{\,1}$ and M.A. Rodr\'iguez$^{\,2}$}

\Address{$^{1}$ Dipartimento di Matematica, Universit\`a degli Studi di
Milano, via Saldini 50, 20133 Milano (Italy)  {\it and} SMRI, 00058 Santa Marinella (Italy); {\tt giuseppe.gaeta@unimi.it} \\[2mm]
$^{2}$ Departamento de F\'{\i}sica Te\'orica,
Universidad Complutense de Madrid, Pza.~de las Ciencias 1, 28040 Madrid (Spain); {\tt rodrigue@ucm.es} }

\Date{Received July 7, 2022; Accepted September 20, 2022}

\setcounter{equation}{0}

\begin{abstract}
\noindent
It is known that knowledge of a symmetry of a scalar Ito stochastic differential equations leads, thanks to the Kozlov substitution, to its integration. In the present paper we provide a classification of scalar autonomous Ito stochastic differential equations with simple noise possessing symmetries; here ``simple noise'' means the noise coefficient is of the form $\s (x,t) = s x^k$, with $s$ and $k$ real constants. Such equations can be taken to a standard form via a well known transformation; for such standard forms we also provide the integration of the symmetric equations. Our work extends previous classifications in that it also considers recently introduced types of symmetries, in particular standard random symmetries, not considered in those.
\end{abstract}

\label{firstpage}


\tableofcontents

\section{Introduction}

When studying nonlinear deterministic differential equations, ODEs and PDEs alike, the use of the symmetry properties of these \cite{\symmref} is an invaluable tool. Actually, what is nowadays known as Lie theory, or the theory of Lie groups and algebras, was first created by Sophus Lie precisely as a tool to solve nonlinear equations. It turned out that treating concrete problems lead to rather complex (albeit in principle elementary) computations, so that the Lie method for studying nonlinear equations was not extensively used until technology provided computers and algebraic manipulation programs. Nowadays all the needed computations can be performed in an automated way, and the complexity of these is not any more an obstacle to application of Lie theory.

In recent years, it was realized that Lie theory can be of great help also in tackling \emph{stochastic} differential equations (SDE) \cite{\sderef}. This presents several problems compared with the deterministic case; the major ones are:
\begin{itemize}
\item[(a)] The lack of a \emph{geometric} theory (which in the deterministic case is provided by the formalism of jet bundles, developed by Cartan and Ehresmann), so that one has only the option of an \emph{algebraic} approach;
\item[(b)] The difficulty in identifying what should be the set of allowed transformations; these actually differ depending on what is the goal of the theory.
\end{itemize}

Here we will mainly be concerned with the problem of \emph{exactly integrating an  Ito stochastic differential equation} thanks to its \emph{symmetry properties} \cite{GRQ1,GRQ2,Unal}. (This will require to clarify point (b) above in our context.)

There is a constructive approach to this problem, due to R. Kozlov \cite{\kozref}: once a symmetry of the given SDE is determined, one performs a change of variables taking the symmetry to its standard form, and in these ``adapted'' coordinates the SDE is promptly integrated.

It should be stressed that this is meant in the stochastic sense: that is, if one knows the initial conditions \emph{and} the realization of the driving stochastic process, then one can describe exactly the realization of the stochastic process defined by the SDE.

The general theory has been developed in a number of publications; see e.g. \cite{\kozref,GS17,GL1,GL2,GSW,Glogistic,GSclass,GOU,GKS} or the review paper \cite{GGPR}; we will recall some basic facts in Section \ref{sec:symmIto} below, referring to the literature (see e.g. the papers quoted above) for details.

An example of application of this approach is provided  by the symmetry integration of the one-dimensional \emph{stochastic logistic equation} with so called \emph{environmental noise} \cite{Glogistic}; in mathematical terms, this corresponds to \emph{multiplicative noise}. In a recent paper \cite{GSclass}, scalar Ito equations with multiplicative noise were classified from the point of view of their symmetry properties; it turns out that the list of symmetric equations is rather limited, but not exceedingly poor.

The purpose of this paper is to extend the symmetry classification of scalar Ito equations beyond the case of multiplicative noise.

More precisely, we will consider the general case of noise coefficient $\s (x,t)$ any smooth function, but will actually consider in more detail -- and obtain more explicit results for -- the case of \emph{time-autonomous} noise coefficient $\s = \s (x)$, and in particular the case where the noise depends in a simple way on the stochastic process $x(t)$ described by the equation; we will therefore speak of \emph{simple noise}.

As well known \cite{\sderef}, a general scalar Ito equation is written as
\beql{eq:Ito} d x \ = \ f(x,t) \, dt \ + \ \s (x,t) \, d w \ , \eeq
where $w = w(t)$ is a Wiener process, also referred to as the driving process.

We will refer to $f(x,t)$ as the \emph{drift term}, or \emph{drift coefficient}, and to $\s (x,t)$ as the \emph{noise term}, or \emph{noise coefficient}; we will always understand that $\s$ is not identically zero (or we would have a non-stochastic equation). The Ito equation is \emph{autonomous} -- by this we will always mean \emph{time-autonomous} -- if both the noise term and the drift term are independent of $t$; we will focus mainly on this case.

We say that we are in the presence of \emph{simple noise} in the case where
\beql{eq:sn} \s (x,t) \ = \ s \ x^k \ , \eeq
with $s$ and $k >0$ real constants; we note that $k$ is not necessarily an integer.

We stress that studying in more detail the autonomous case is not only due to the interest of this case in applications: a theorem by Kozlov (see Theorem 3.1 in \cite{Koz2}) states that if a general Ito equation admits a symmetry, then there is a change of variables -- also provided by the theorem -- mapping it into a time-autonomous equation.

It should also be mentioned that a symmetry classification of scalar Ito equations was provided by Kozlov \cite{Koz2}; our initial motivation to go again over this computation was manifold: \begin{itemize}
\item On the one hand, this classification is not completely explicit, so that using it requires some effort;
\item On the other hand, the Kozlov classification is in a way exceedingly fine: in fact, it was shown by Kozlov himself that only symmetries not acting on time are useful for integrating the Ito equations; we thus wanted to provide a rougher classification looking only at symmetries of this type.
\item Moreover, in many applications one is interested only in the case of \emph{simple noise}, see above; restricting to this case allows a more explicit classification, as we will see.
\item Finally, and more substantially, after the work by Kozlov different kind of symmetries (random standard symmetries and W-symmetries, see below) which were not considered at the time have been introduced, and we want to have a classification covering these as well.
\end{itemize}

The relation of our classification to Kozlov's one is briefly discussed in Appendix \ref{app:Kcompare}.

It is maybe worth noting that we discuss general Ito equations; the special case of equations with a variational origin, of obvious physical interest and discussed (also from the symmetry point of view) in the literature (see e.g. \cite{Misvar,Yas,Zam1,Zam2,CruZam,TZ}) will not be specifically dealt with. Similarly, the approach to symmetry of Ito equations through Girsanov theory \cite{ARZ,DVMU1,DVMU2} lies beyond the limits of our work.


\section{Symmetries of Ito equations}
\label{sec:symmIto}

We start by recalling some basic facts concerning symmetries of Ito equation; we will confine ourselves to the case of scalar equations, and will refer to these as written in the form \eqref{eq:Ito}.

\subsection{Admissible symmetries}

We are only interested in Lie-point (as opposed to discrete) symmetries; moreover we will only consider symmetries which do \emph{not} transform the time variable $t$ into a stochastic process\footnote{For a more general approach taking this possibility into account, see e.g. \cite{DVMU1,DVMU2}.}. Thus the most general form of the considered symmetry vector fields will be
\beql{eq:Ygen} Y \ = \ \vphi (x,t;w) \, \pa_x \ + \ \tau (t) \, \pa_t \ + \ h(x,t;w) \, \pa_w \ . \eeq
Note that this allows for a reparametrization of time; for this to be a proper one, we should require that $\tau $ is a monotone (say, growing) function of $t$, i.e. $\tau' (t) > 0$. This option will be of little interest in the following, and later on we will just set \beq \tau \ = \ 0 \ . \eeq

The last term, $h(x,t;w) \pa_w$, corresponds to an action on the Wiener process $w = w(t)$. As we want that the Ito equation \eqref{eq:Ito} is transformed into an Ito equation (actually, the same), the transformed process should still be a Wiener process. This requires that $h(x,t;w)$ only depends on $w$, and moreover it can only give a rescaling -- which, if we require the equation to be invariant, can be reabsorbed by a change on the noise coefficient $\s (x,t)$ -- and this implies that
\beql{eq:h} h(x,t;w) \ = \ r \, w \ . \eeq
We refer to \cite{GS17,GSclass,GKS} for further discussion on this matter.

It will be convenient to have some nomenclature for the different types of symmetries; these correspond of course to the case where the Ito equation is invariant under $Y$, and we will see the condition for this to be the case in a moment.

\begin{itemize}
\item When $h \not= 0$ (i.e., with the restricted form set in \eqref{eq:h}, when $r\not= 0$), we say that we have a \emph{W-symmetry}\footnote{Sometimes this is also called a \emph{proper} W-symmetry, to emphasize that standard symmetries (see in a moment) are a special case of W-symmetries. Here we consider as W-symmetries just proper W-symmetries, for ease of language.};
\item if $h = 0$ (i.e., see \eqref{eq:h}, when $r=0$) we have a \emph{standard symmetry}.
\item Within the class of standard symmetries ($r=0$), we will distinguish between the case where $\vphi$ depends effectively on $w$, in which case we speak of \emph{random symmetries};
\item and that of standard symmetries ($r=0$) where $\vphi$ is independent of $w$, i.e. $\vphi_w = 0$, $\vphi = \vphi (x,t)$, in which case we speak of \emph{deterministic symmetries}.
\end{itemize}

\medskip\noindent
{\bf Remark 1.} It should be stressed that applying the Kozlov change of variables related to a standard symmetries we are guaranteed to map an Ito equation into an Ito equation; on the other hand, if this is done with a W-symmetry, the equation we obtain is not guaranteed in general to be an Ito equation. Thus, albeit it may happen that this equation can be solved exactly, we do not have a general theory for this setting, i.e. for the case of W-symmetries. A similar \emph{caveat} holds for random standard symmetries; see Remark 4 below in this regard.  \EOR

\subsection{Determining equations}

We will thus be considering vector fields of the form ($r$ a constant)
\beql{eq:Y} Y \ = \ \vphi (x,t;w) \, \pa_x \ + \ r \, w \, \pa_w \ . \eeq

Such a vector field is a symmetry of the Ito equations \eqref{eq:Ito} if and only if its coefficients satisfy the \emph{determining equations}
\begin{eqnarray}
& & \vphi_t \ + \ f \, \vphi_x \ - \ \vphi \, f_x \ + \ \frac12 \, \Delta (\vphi ) \ = \ 0 \ , \label{eq:deteq1} \\
& & \vphi_w \ + \ \s \, \vphi_x \ - \ \vphi \, \s_x \ - \ r \, \s \ = \ 0 \ ; \label{eq:deteq2} \end{eqnarray}
here and in the following, $\Delta$ is the \emph{Ito Laplacian}, which in this simple scalar case reads
\beql{eq:Delta} \Delta (\phi) \ := \ \phi_{ww} \ + \ 2 \, \s \, \phi_{xw} \ + \ \s^2 \, \phi_{xx} \ . \eeq

We refer e.g. to \cite{GS17,GSclass,GGPR} for the derivation of these determining equations and for a discussion of their properties.

Here we just note that \eqref{eq:deteq1} is a second order PDE, depending on both the $f$ and the $\s$ coefficient in the Ito equation (the dependence on the $\s$ coefficient being through the Ito Laplacian), while \eqref{eq:deteq2} is a first order PDE, depending only on the noise coefficient $\s$, but not on the drift coefficient $f$ (see however Remark 2 below). It is thus not surprising that, in general, it will be easier to tackle \eqref{eq:deteq2} than \eqref{eq:deteq1}.

In fact, our general strategy will be to use \eqref{eq:deteq2} to restrict the functional form of $\vphi (x,t;w)$ before tackling the more complex equation \eqref{eq:deteq1}.

\medskip\noindent
{\bf Remark 2.} The structure of the system made of eqs. \eqref{eq:deteq1}, \eqref{eq:deteq2} is that of a set of \emph{linear, non homogeneous} equations for $\vphi$. Thus we know that the most general solution for this will be made by the superposition of a particular solution for the full system -- which for ease of reference we will also denote as $\omega (x,t,w)$ -- and of the most general solution for the associated homogeneous system. The latter is just the system of determining equations for standard symmetries of the Ito equation under study. In other words, when we look for the most general proper W-symmetries of an Ito equation, these will also include an additive term corresponding to a general standard symmetry. In order to avoid unneeded notational complications, when looking for proper W-symmetries we will just set to zero the standard part. In the cases where some ambiguity could arise, we will refer to proper W-symmetries with zero standard part as \emph{strict proper W-symmetries}. \EOR

\subsection{Determining equations in first order form}
\label{sec:firstorder}

We can use \eqref{eq:deteq2} and its differential consequences to simplify \eqref{eq:deteq1}. In fact, \eqref{eq:deteq2} yields $\vphi_w = \vphi \s_x - \s \vphi_x + r \s$; differentiating this w.r.t. $x$ and $w$ we get respectively
\begin{eqnarray*}
\vphi_{xw} &=& \s_{xx} \, \vphi \ + \ \s_x \, \vphi_x \ - \ \s_x \, \vphi_x \ - \ \s \, \vphi_{xx} \ + \ r \, \s_x \\
&=& \s_{xx} \, \vphi \ - \ \s \, \vphi_{xx} \ + \ r \, \s_x \ , \\
\vphi_{ww} &=& \s_x \, \vphi_w \ - \ \s \, \vphi_{xw} \\
&=& \s_x \( \vphi \s_x - \s \vphi_x + r \s \) \ - \ \s \, \( \s_{xx} \, \vphi \ - \ \s \, \vphi_{xx} \ + \ r \, \s_x \) \\
&=& \( \s_x^2 \, - \, \s \, \s_{xx} \) \, \vphi \ - \ \s \, \s_x \, \vphi_x \ + \ \s^2 \, \vphi_{xx} \ .
 \end{eqnarray*}
Substituting these into the expression for the Ito Laplacian \eqref{eq:Delta} we get
\begin{eqnarray}
\De (\vphi ) & := & \vphi_{ww} \ + \ 2 \, \s \, \vphi_{xw} \ + \ \s^2 \, \vphi_{xx} \nonumber \\ & = &
 \( \s_x^2 \, + \, \s \, \s_{xx} \) \, \vphi \ - \ \s \, \s_x \, \vphi_x  \ + \ 2 \, r \, \s \, \s_x  \ . \end{eqnarray}
It follows that \eqref{eq:deteq1} is now written as a first order PDE for $\vphi$.

More precisely, we get
\beq \vphi_t \ + \ \( f \, - \, \frac12 \, \s \, \s_x \) \, \vphi_x \ - \ \( f \, - \, \frac12 \, \s \, \s_x \)_x \, \vphi \ + \ r \, \s \, \s_x \ = \ 0 \ . \eeq
Introducing the modified drift
\beql{eq:f_frak} b \ := \  f \, - \, \frac12 \, \s \, \s_x \ , \eeq
this is also written as
\beql{eq:R2t} \vphi_t \ + \ b \,  \vphi_x \ - \ b_x \, \vphi \ + \ r \, \s \, \s_x \ = \ 0 \ . \eeq
We have thus shown that the system \eqref{eq:deteq1}, \eqref{eq:deteq2} is equivalent to the system \eqref{eq:R2t}, \eqref{eq:deteq2}. Note that $b = b(x,t)$ is the Stratonovich drift for the Ito equation, and these are the determining equations for the Stratonovich equation associated to our Ito equation \cite{Koz18b,GL1,GGPR}.

Alternatively, we can substitute for $r \s$ in \eqref{eq:R2t} according to \eqref{eq:deteq2}; this yields
\beql{eq:R2tb} \vphi_t \ + \ \s_x \, \vphi_w \ + \ \( b \ + \ \s \, \s_x \) \, \vphi_x \ - \ \( b_x \ + \ \s_x^2 \) \, \vphi  \ = \ 0 \ . \eeq

We are thus reduced to the study of a system of two first order linear equations, i.e. \eqref{eq:R2t}, \eqref{eq:deteq2} in the general case, or alternatively the system \eqref{eq:R2tb}, \eqref{eq:deteq2}. One should not be too optimistic about solving this: in fact, our forthcoming discussion will show that in general it does not admit any solution.

\medskip\noindent
{\bf Remark 3.} The relation between symmetries of an Ito and of the associated Stratonovich equation has been investigated in a number of papers \cite{GL1,GL2,GSW,Koz18a,Koz18b,Unal}. As for the relation between \emph{W-symmetries} of an Ito and of the associated Stratonovich equation, this was studied in \cite{GSW}. In particular, Theorem 1 therein, when applied to the present case of scalar equations, states that W-symmetries of an Ito equation are also symmetries of the associated Stratonovich equation (and vice versa) if and only if the diffusion coefficient $\s (x,t)$  is spatially constant, $\s_x = 0$. It is a simple consequence of this result that, as stated in Theorem 2 of the same paper (again restricting it to the case of scalar equations), W-symmetries of an Ito equation are preserved under an admissible change of variables if and only if the diffusion coefficient is spatially constant. \EOR

\subsection{Compatibility conditions}
\label{sec:compatibility}

The determining equations in first order form, \eqref{eq:R2t} and \eqref{eq:deteq2}, also provide some compatibility conditions between $b$, $\s$ and $\vphi$ (or equivalently between $f$, $\s$ and $\vphi$). In fact, differentiating \eqref{eq:R2t} w.r.t. $w$ and \eqref{eq:deteq2} w.r.t. $t$, recalling that neither $b$ nor $\s$ depend on $w$, and taking the difference of these differentiated equations, we get
\beq ( b  \ - \ \s ) \ \vphi_{xw} \ - \ (b_x \ - \ \s_x ) \ \vphi_w \ = \ 0 \ . \eeq This reduces to a trivial identity for $\vphi_w = 0$. For $\vphi_w \not= 0$, we have that $\vphi$ is necessarily of the form
\beq \vphi (x,t;w) \ = \ \[ b(x,t) \ - \ \s (x,t) \] \ Q (t,w) \ , \eeq
with $Q$ an arbitrary function.

Note that in the simplest case we are considering later on, i.e. for $\s = \s (x)$ and $f = f(x)$ -- which together also entail $b = b(x)$ -- we have a separation of variables in $\vphi$, i.e. this is of the form
\beq \vphi (x,t;w) \ = \ \[ b(x) \ - \ \s (x) \] \ Q (t,w) \ , \eeq
with $Q$ an arbitrary function encoding all the dependencies of $\vphi$ on both $t$ and $w$.

In the case $r=0$, eq. \eqref{eq:deteq2} reads then
\beq \frac{Q_w}{Q} \ = \ \frac{b \, \s_x \ - \ b_x \, \s}{b \ - \ \s} \ ; \eeq as the l.h.s. of this is a function of $t$ and $w$, while the r.h.s. is a function of $x$ alone, both must be actually constants. This means not only the functional form of $Q$ is further constrained,
\beql{eq:RR_Q} Q(t,w) \ = \ \eta (t) \ e^{k w} \ ; \eeq but also that there is a compatibility condition between $b$ and $\s$ (i.e. between $f$ and $\s$) to admit symmetries. In fact, we need
\beq b \, \s' \ - \ b' \, \s \ = \ k \ \( b \ - \ \s \) \eeq for some constant $k$.
This can be expressed as $b$ being a solution to
\beql{eq:RR_B} b' \ = \ \frac{k \, (\s - b) \ + \ b \, \s'}{\s} \ . \eeq
Plugging \eqref{eq:RR_Q} and \eqref{eq:RR_B} into \eqref{eq:R2t} (and recalling we assumed $r=0$) we get
\beq e^{k w} \ \( b(x) \ - \ \s (x) \) \ \( \eta' (t) \ - \ k \, \eta (t) \) \ = \ 0 \ . \eeq
This means that, unless we are in the very special case where $b = \s$, i.e. where
$$ f(x) \ = \ \s (x) \ + \ \frac12 \, \s (x) \, \s' (x) \ , $$
necessarily we have
$$ \eta (t) \ = \ e^{k t} \ ; $$ hence $Q$ is of the form
\beq Q(t,w) \ = \ q \ \exp [ k (t+w) ] \ , \eeq
with $q$ and $k$ constants, as follows at once from \eqref{eq:RR_Q}.

\section{Symmetry and integration of the Ito equation}
\label{sec:geom}

Before going into our classification task, it is convenient to discuss what will be the use of symmetries if we find some. This will also provide a justification to our choice of restricting the form of the considered vector fields.

As for the case of deterministic ODEs -- and as anticipated in the Introduction -- knowledge of a \emph{standard} Lie-point symmetry allows to determine constructively a change of variables
$$ (x,t;w) \ \to \ (y,t;w) $$
upon which the resulting equation for $y(t)$ -- which is of course different from the original one, albeit equivalent to it -- is readily integrated, providing an explicit realization $y(t)$ for any realization of the driving Wiener process $w(t)$; inverting the change of variables this provides an explicit realization $x(t)$ for the process described by the original Ito equation.

\medskip\noindent
{\bf Remark 4.} Note that in this formulation the equation for $y(t)$ does not need to be of Ito type; however, stochastic equations of more general type do lack a solid mathematical foundation, so it is preferable to consider only cases leading to Ito type SDE for $y(t)$. This point is discussed in detail e.g. in \cite{GOU} (see Remark 12 therein), to which we refer for further detail. In the case of \emph{deterministic standard symmetries} we are guaranteed that the new random variable $y(t)$ obtained applying the Kozlov transformation obeys an Ito equation, so that this is a substantial reason to study this standard symmetry case in more detail.
\EOR
\bigskip

It should be mentioned that albeit in dealing with W-symmetries the integration procedure is conceptually equivalent (being based on symmetries and transformation to adapted variables), its practical realization is somewhat more delicate, in that one should also act on the Wiener variables $w^i$; this might present some subtleties. This feature will not be relevant here, as we will found that W-symmetries are present only in a trivial way in our problem (see Appendix \ref{app:noW}); thus the reader is referred to \cite{GSW}, see in particular Section VIII therein, for a discussion of this point.

Assume now that a \emph{standard symmetry}
\beql{eq:Ysimp} Y \ = \ \vphi (x,t;w) \, \pa_x \eeq for equation \eqref{eq:Ito} has been determined. We perform then the \emph{Kozlov change of variables} \cite{Koz1,Koz2,Koz3}
\beql{eq:Koz} y \ = \ \Phi (x,t;w) \ := \ \int \frac{1}{\vphi (x,t;w) } \ d x \ . \eeq

The equation for the new variable $y = y(t,w)$ is then in the form (see Appendix \ref{app:Koz} for details)
\beql{eq:ItoS} d y \ = \  F(t,w ) \, dt \ + \ S(t,w ) \, d w \ ; \eeq
therefore
\beql{eq:Ydef} y(t) \ = \ y (t_0) \ + \ \int_{t_0}^t F[\tau , w(\tau)]  \, d \tau \ + \ \int_{t_0}^t S [\tau , w (\tau)] \, d w (\tau ) \ , \eeq
which provides the solution in closed form: for any realization of the driving Wiener process $w (t)$, we have explicitly the corresponding realization of the process $y (t)$\footnote{See e.g. the numerical experiments reported in \cite{Glogistic} for a pictorial illustration of this discussion.}. Note that if $\vphi$ is a \emph{deterministic standard symmetry}, i.e. $\vphi = \vphi (x,t)$, then in \eqref{eq:ItoS} we have $F = F(t)$, $S = S(t)$, and \eqref{eq:Ydef} reduces to
\beql{eq:Ydefdeter} y(t) \ = \ y (t_0) \ + \ \int_{t_0}^t F(\tau)  \, d \tau \ + \ \int_{t_0}^t S (\tau) \, d w (\tau ) \ . \eeq
On the other hand, for a general $\vphi$, i.e. if this also depends on $w$, we are not guaranteed \eqref{eq:ItoS} is in Ito form. See also \cite{GOU}.

\medskip\noindent
{\bf Remark 5.} It may be worth writing down the explicit relation between the original drift and noise coefficients $f(x,t)$ and $\s (x,t)$ on the one hand, and the transformed ones $F(t,w)$ and $S(t,w)$ on the other. These are
\begin{eqnarray}
F(t,w) &=& \frac{f}{\vphi} \ - \ \frac12 \, \sigma^2 \, \frac{\vphi_x}{\vphi^2} \  - \ \int \frac{\vphi_t}{\vphi^2} \ d x  \ , \nonumber  \\
S(t,w) &=& \frac{\sigma}{\varphi} \ - \ \int \frac{\vphi_w}{\vphi^2} dx \ . \end{eqnarray}
We refer to Appendix \ref{app:Koz} or to \cite{\kozref} for the derivation of these formulas. \EOR

\section{Reduction of Ito equations to standard form}
\label{sec:standard}

When dealing with Ito equations, it is convenient to reduce them to a \emph{standard form}; this procedure will be discussed in this section and here we will always assume $\s (x,t) \not= 0$.

Consider the function
\beql{eq:g} g(x,t) \ := \ \int \frac{1}{\s (x,t)} \ d x \ , \eeq
i.e. a primitive of $\s$ w.r.t. the variable $x$; this is always defined up to a function of $t$ alone, which we will take to be zero.

In the following, it will be convenient to also define the function
\beql{eq:varrho} \varrho (x,t) \ := \ - \ \frac{\pa}{\pa t} g (x,t) \ = \
\int  \frac{\s_t (x,t)}{\s^2 (x,t)} \ d x \ . \eeq

Note that for $\s$ autonomous, $\s = \s (x)$, the function $g$ will also be a function of $x$ alone, $g = g (x)$; correspondingly, $\varrho (x,t)$ is identically zero if $\s$ does not depend on time.

We can then consider the change of variable
\beql{eq:yg} y \ = \ g(x,t) \ . \eeq
By Ito rule, and with $\Lap$ the Ito Laplacian \eqref{eq:Delta}, we have
\begin{eqnarray*}
d y &=& \frac{\pa g}{\pa x} \, d x \ + \ \frac{\pa g }{\pa t} \, d t \ + \ \frac12 \, \Lap (g) \, d t \\
&=& \frac{1}{\s (x,t) } \, \[ f(x,t) \, dt \ + \ \s (x,t) \, d w\]  \ - \ \( \int \frac{\s_t (x,t)}{\s^2 (x,t)} \, d x \) \, d t \\
& & \ \ \ \ - \ \frac12 \, \s^2 (x,t) \, \( \frac{\s_x (x,t)}{\s^2 (x,t)} \) \, dt \\
&=& \[ \frac{f (x,t)}{\s (x,t) } \ - \ \frac12 \, \s_x (x,t) \ - \ \varrho (x,t) \] \ d t \ + \ d w \\
&:=& F[x(y,t),t] \, dt \ + \ d w \ . \end{eqnarray*}

In the last line, we should express $x$ as a function of $y$ and $t$ inverting the change of variables $x \to y = g(x,t)$; in the following we will denote this by \beql{eq:xxi} x \ = \ \xi (y,t) \ . \eeq

In this way, i.e.  through the change of variable \eqref{eq:yg}, \eqref{eq:g}, any Ito equation \eqref{eq:Ito} is changed into an Ito equation with unit noise term
\beq d y \ = \ F (y,t) \, dt \ + \ d w \ , \eeq
and the drift coefficient $F (y,t)$ is defined in terms of the original drift and noise coefficients (and of our shorthand definition \eqref{eq:varrho}) as
\beql{eq:Fgenyt} F(y,t) \ = \ \[ \frac{f (x,t)}{\s (x,t) } \ - \ \frac12 \, \s_x (x,t) \ - \ \varrho (x,t) \]_{x = \xi (y,t)} \ . \eeq

\medskip\noindent
{\bf Remark 6.} Note that if the original Ito equation is time-autonomous, i.e. $f=f(x)$ and $\s = \s (x)$, then the transformed Ito equation is also time-autonomous, i.e. $f = F(y)$; more precisely, in this case \eqref{eq:yg} and \eqref{eq:xxi} yield $y = g(x)$, $x = \xi (y)$, and we get
\beql{eq:Fgenaut} F(y) \ = \ \[ \frac{f (x)}{\s (x) } \ - \ \frac12 \, \s_x (x)  \]_{x = \xi (y)} \ . \eeq

\medskip\noindent
{\bf Remark 7.} One may be tempted to use the reduction to standard form also for integrable equations, so to have a simple Brownian integral. In fact, consider an equation of the form
\beq d x \ = \ f(t) \, d t \ + \ \s (t) \, d w \ ; \eeq
this is integrated, yielding
\beq x(t) \ = \ x (t_0) \ + \ \int_{t_0}^t f(\tau) \ d \tau \ + \ \int_{t_0}^t \s (\tau) \, d w (\tau) \ . \eeq However, determining the function $x(t)$ for a given realization of the Wiener process $w(t)$ is in practice non trivial.

Passing to the variable
$$ y \ = \ g (x,t) \ = \ \int \frac{1}{\s (t) } d x \ = \ \frac{x}{\s (t)} $$ we have (recalling that $\Delta (g)=0$, as $y=g(x,t)$ is a linear function of $x$)
\begin{eqnarray*}
dy &=& g_x \, d x \ + \ g_t \, dt \\ &=& \frac{1}{\s(t)} \( f(t) \, dt \ + \ \s (t) \, dw \) \ - \ \( \frac{\s_t (t)}{\s^2 (t)} \) \ x \ d t \\
&=& \[ \( \frac{f(t)}{\s(t)} \) \ - \ \( \frac{\s_t (t)}{\s (t)} \) \, y \] \, d t \ + \ d w \\
&:=& \[ a(t) \ + \ b(t) \ y \] \ d t \ + \ d w \ . \end{eqnarray*}
Thus we have transformed the original equation -- which was independent of $x$ and hence immediately integrable -- into an equation which is not \emph{immediately} integrable. Needless to say, the equation is still integrable, albeit showing this will require some further work. See Section \ref{sec:intB}.
\EOR

\medskip\noindent
{\bf Remark 8.} A special case is provided by $\s (x,t) = s$; this is reduced to the case of unit noise coefficient by just $y = x/s$ and $x = s y$, see \eqref{eq:DCC} and \eqref{eq:ICC}, and in this case we get
\beq F(y,t) \ = \ \frac{1}{s} \ f ( s y , t) \eeq
by just applying \eqref{eq:Fgenyt}. \EOR

\section{The case of constant noise}
\label{sec:const}

We want to discuss right away the case of \emph{constant noise}. This is definitely relevant in applications, and -- albeit it was at the basis of the Kozlov classification \cite{Koz2} -- was not considered in previous papers dealing also with random and W-symmetries \cite{Glogistic,GSclass}, so we should look at it.

Moreover, and more importantly, the discussion of Section \ref{sec:standard} above shows that this case -- and actually that with constant unit noise --   will provide the skeleton for a classification in the general case.

In view of the relevance for applications of the general constant noise case, we will first consider the case $\s (x,t) = s$, and only later on restrict to the case $s = 1$; we could of course also consider just the case $s=1$ and use Remark 8 above.

For $\s (x,t) = s$, the second determining equation \eqref{eq:deteq2} with $r=0$ reads simply
\beq \vphi_w \ + \ s \, \vphi_x \ = \ 0 \ ; \eeq
this is solved immediately to yield
\beql{eq:vphi0} \vphi (x,t;w) \ = \ \psi ( u , t) \ ; \ \ \ u \ := \ x \ - \ s \,  w \ . \eeq

The situation is only slightly more complex for general $r$; in fact, in this case the equation  \eqref{eq:deteq2} reads
\beq \vphi_w \ + \ s \, \vphi_x \ = \  r \, s  \ ; \eeq
this is also solved immediately, now to yield (with $u$ as above)
\beql{eq:vphir} \vphi (x,t;w) \ = \ r \, x \ + \ \psi ( u , t) \ . \eeq
Note it is legitimate to use the same symbol $\psi$ as in \eqref{eq:vphi0} above, since in this way for $r=0$ we are reduced indeed to \eqref{eq:vphi0} itself.

Note also that $\s (x,t) = s$ and $u = x - s w$ entail that
\beq \Delta [ \psi (u,t) ] \ = \ 0 \ . \eeq

The first determining equation \eqref{eq:deteq1} is then
\beql{eq:vpsir} \psi_t \ + \ f \, \psi_u \ - \ \psi \, f_x \ + \ r \ \( f \ - \ x \, f_x \) \ = \ 0 \ ; \eeq
in the case $r=0$ this is, of course, just
\beql{eq:vpsi0} \psi_t \ + \ f \, \psi_u \ - \ \psi \, f_x \ = \ 0 \ ; \eeq

We will now first treat the special cases of $f(x,t)$ independent or linearly dependent on $x$, and then deal with the general one excluding these special, and in some respect degenerate, ones. (More precisely, these are degenerate in that they give $f_{xx} (x,t) = 0$; see the discussion below.)

\subsection{Case A. Drift independent of $x$}

Let us first consider the case where
\beq f(x,t) \ = \ h (t)  \eeq
(possibly with $h(t) = c$ or $h(t) = 0$); we will denote by $H(t)$ a primitive of $h(t)$,
\beql{eq:hprim} H(t) \ = \ \int h(t) \ dt \ . \eeq

Note that this case is of little interest, in that it is trivially integrable to give
$$ x(t) \ = \ x(t_0) \ + \ \int_{t_0}^t h (\tau) \ d \tau \ + \ s \ [ w(t) \, - \, w (t_0) ] \ , $$ but we will analyze its symmetries for the sake of completeness.

Now the equation \eqref{eq:vpsir} reads
\beq \psi_t (u,t) \ + \ \[ r \ + \ \psi_u (u,t) \] \ h(t) \ = \ 0 \ ; \eeq
this is readily solved, e.g. by the method of characteristics, yielding with the notation \eqref{eq:hprim}
\beq \psi (u,t) \ = \ P \[ u - H(t) \] \ - \ r \, H(t)  \eeq
where $P$ is an arbitrary function of its argument.
As for the corresponding symmetry vector field $Y$, see \eqref{eq:Y}, this is identified by
\beq \vphi (x,t;w) \ = \ P [ x \, - \, s \, w \, - \, H (t) ] \ + \ r \ \[ x \, - \, H (t) \] \ . \eeq
In the case of standard symmetries, $r=0$, this just reduces to
\beq \vphi (x,t;w)  \ = \ P [ x \, - \, s \, w \, - \, H (t) ] \ . \eeq

\subsection{Case B. Drift linear in $x$}

We consider next the case where
\beql{eq:fB} f(x,t) \ = \ h(t) \ + \ k(t) \, x \ ; \eeq
we assume $k(t) \not= 0$, or we would fall back into case A seen above. Similarly to what we have done before, we denote a primitive of $k$ as
\beql{eq:kprim} K(t) \ = \ \int k(t) \ dt \ . \eeq

The equation \eqref{eq:vpsir} reads now
\beq \psi_t \ + \ (h + k x) (r + \psi_u ) \ - \ k \ (r x + \psi ) \ = \ 0 \ ; \eeq or, with a little simplification,
\beql{eq:vpsirB} \psi_t \ + \ h \, \psi_u \ - \ k \, \psi \ + \ r \, h \ + \ \( k \, \psi_u \) \, x \  = \ 0 \ .
\end{equation}

Obviously \eqref{eq:vpsi0} is obtained from this by setting $r=0$.

As $h = h(t)$ and $\psi = \psi (u,t)$ do not depend on $x$, the $x$-dependencies in \eqref{eq:vpsirB} are explicit. Looking at the coefficient of $x$ in \eqref{eq:vpsirB}, for any value of $r$ we have
\beq k(t) \ \psi_u (u,t) \ = \ 0 \ . \eeq
As by assumption $k(t) \not= 0$, this requires $\psi_u = 0$, i.e.
\beq \psi (u,t) \ = \ \eta (t) \ . \eeq
The equation \eqref{eq:vpsirB} is thus reduced to
\beql{eq:etaB} \eta' (t) \ = \ k(t) \ \eta (t) \ - \ r \, h(t) \ , \eeq
while in the case $r=0$ we simply have
\beql{eq:etaB0} \eta' (t) \ = \ k(t) \ \eta (t) \ . \eeq
The latter is promptly solved by
\beq \eta (t) \ = \ A \ \exp [ K(t)] \ ; \eeq that is, up to an inessential multiplicative constant we have only one standard symmetry, identified by
\beq \vphi (x,t;w) \ = \ \exp [ K(t) ] \ = \ \exp \[ \int k(t) \, dt \] \ . \eeq

As for W-symmetries, i.e. for the case $r \not= 0$, eq.\eqref{eq:etaB} is solved by
\beq \eta (t) \ = \ A \ e^{K(t)} \ - \ r \, e^{K(t)} \ \int h(t) \ e^{- K(t)} \ d t \ , \eeq
yielding
\beq \vphi (x,t;w) \ = \ A \ e^{K(t)} \ + \ r \, x \ - \  r \, e^{K(t)} \ \int_0^t e^{- K (\tau) } \ h(\tau) \ d \tau \ . \eeq

\subsection{Case C. General drift}

We will assume from now on that the first two $x$-derivatives of $f(x,t)$ are nonzero, i.e. $f_{xx} \not= 0$ (which also implies $f_x \not= 0$), having treated separately the case of $f$ independent of $x$ or linear in $x$.

Our equation \eqref{eq:vpsir} depends on $x$ only through an explicit term $r x$ and through $f$ and $f_x$; as the series expansions of $f$ and $f_x$ are independent unless $f$ is an exponential in $x$, we expect to get standard symmetries only for an exponential $f$, possibly with a slight modification for W-symmetries due to the presence of the term $rx$. This will indeed be the case.

Let us first of all note that if there is a $\psi$ satisfying this equation, it must actually depend on $u$. In fact, assume $\psi (u,t) = \eta (t)$; then \eqref{eq:vpsir} reads
$$ \eta' (t) \ + \ r \, f(x,t) \ - \ [ r \, x \ + \ \eta (t) ] \ f_x (x,t) \ = \ 0 \ . $$ Differentiating this w.r.t. $x$ we get
$$ [ r \, x \ + \ \eta (t) ] \ f_{xx} (x,t) \ = \ 0 \ . $$
But we assumed $f_{xx} (x,t) \not= 0$, so it should be
$$ r \, x \ + \ \eta (t) \ = \ 0 \ , $$ which requires to have \emph{both} $r=0$ \emph{and} $\eta (t) = 0$. Thus we conclude that indeed a nonzero $\psi (u,t)$ must depend effectively on $u$.

Actually, as we show in a moment, it has to satisfy $\psi_{uu} \not= 0$. In fact, consider again \eqref{eq:vpsir} and differentiate it once w.r.t. $u$ and once w.r.t. $x$; this yields
\beql{eq:fpsiu0} f_x  \ \psi_{uu}  \ - \ f_{xx} \ \psi_u \ = \ 0 \ . \eeq
We note that if $\psi_{uu}  = 0$, then the equation reduces to $f_{xx} \psi_u = 0$; but having assumed $f_{xx} \not= 0$, this would require $\psi_u = 0$, hence $\psi (u,t) = \eta (t)$, and we have seen above this implies $\psi = 0$; so we can exclude the case $\psi_{uu}=0$ as well.

We like to write \eqref{eq:fpsiu0}, making use of $f_{xx} \not= 0$ and $\psi_u \not= 0$, in the form
\beq \frac{\psi_{uu}}{\psi_u} \ = \ \frac{f_{xx}}{f_x} \ . \eeq Here the l.h.s. is a function of $u$ and $t$, while the r.h.s. is a function of $x$ and $t$. Thus the equation may hold if and only if
\beq \frac{\psi_{uu} (u,t)}{\psi_u (u,t)} \ = \ B (t) \ = \ \frac{f_{xx} (x,t)}{f_x (x,t)} \eeq for some function $B (t)$. These equations yield
\begin{eqnarray} f(x,t) &=& a_1 (t) \ + \ b_1 (t) \ \exp[B(t) \ x] \ , \\
\psi (u,t) &=& a_2 (t) \ + \ b_2 (t) \ \exp[B(t) \ u] \ ; \end{eqnarray} here $a_i$ and $b_i$ are arbitrary smooth functions of time.

Note that we must require both $b_1 (t) \not= 0$ and $B(t) \not= 0$, or we would be back to the case of $x$-independent drift. We must also require $b_2 (t) \not= 0$, or $\psi$ would not depend effectively on $u$, which we have seen above is instead necessarily the case.

Inserting these expressions into \eqref{eq:vpsir}, we get (omitting the functional dependencies for ease of notation)
\beql{eq:vphiC} e^{x B}  \ \[ r - \( r x + a_2  \) \, B \] \ b_1  \ + \ e^{u B} \ \[ b_2'  \, + \, u \, b_2 \, B' \, + \, a_1 \, b_2 \, B \] \ + \ \( r \, a_1 \, + \, a_2' \) \ = \ 0 \ . \eeq

Differentiating this w.r.t. $x$, we get
\beq e^{x B} \ \(r \, x \ + \ a_2 \) \ b_1 \ B^2 \ = \ 0 \ . \eeq
As noted above, we must assume $b_1 \not=0$, $B \not= 0$. Thus we conclude that $r=0$, which means there are no proper W-symmetries, and $a_2 (t) = 0$.

With these, \eqref{eq:vphiC} reads
\beql{eq:vphiC2} e^{u B} \ \[ b_2' \ + \ a_1 \, b_2 \, B \ + \ u \, b_2 \, B' \] \ ; \eeq as the $u$ dependencies are now explicit, this splits into the two equations
\begin{eqnarray}
& & b_2 (t) \ B' (t) \ = \ 0 \ , \label{eq:eqeqeqeq1} \\
& & b_2' (t) \ + \ a_1 (t) \, b_2 (t) \, B (t) \ = \ 0 \ . \label{eq:eqeqeqeq2} \end{eqnarray}
The first equation requires to have either $b_2 (t) = 0$ or $B(t) = \b$ a constant. But, the case $b_2 (t) = 0$ has been excluded above (it gives $\psi = 0$, i.e. no symmetry). Thus we are left to consider the case $B(t) = \b$, which corresponds to
\beql{eq:fC} f(x,t) \ = \ a_1 (t) \ + \ b_1 (t) \ e^{\b x} \ . \eeq
Now the equation \eqref{eq:eqeqeqeq2} reads
$$ b_2' (t) \ = \ - \ \b \ a_1 (t) \ b_2 (t) \ , $$ and its solution is
\beq b_2 (t) \ = \ c \ \exp \[ - \ \b \ \int a_1 (\tau ) \ d \tau \] \ := \ c \ \exp[ - \b \, A_1 (t) ] \ ; \eeq here we have denoted by $A_1$ a primitive of $a_1$, and $c$ is an arbitrary constant (which may absorb the constant of integration arising from the primitive).

Summarizing, we have concluded that for $f(x,t)$ of general form (that is, not independent of or linearly dependent on $x$), we have symmetries if and only if $f(x,t)$ is of the form \eqref{eq:fC}. In this case the symmetries are standard ones (i.e. $r=0$), and are identified (as usual, up to an inessential multiplicative constant) by
\beql{eq:vphiCfin} \vphi (x,t;w) \ = \ \exp \[ \b \, \( x \, - \, s \, w \ - \ A_1 (t) \) \] \ . \eeq

\subsection{Summary of results for constant noise}

We have thus concluded our detailed discussion of the constant noise (i.e. $\s (x,t) = s$) case; we have completely classified all the drift terms such that the Ito equation has either standard or W-symmetries. Our results are summarized in the following two Propositions (Proposition 1 for proper W-symmetries and Proposition 2 for standard symmetries) and, for standard symmetries in Table \ref{tab:1}; the case of \emph{autonomous} Ito equations is considered in Remark 10 and, for standard symmetries, in Table \ref{tab:2}.

\medskip\noindent
{\bf Proposition 1.} (Proper W-symmetries) {\it An Ito equation \eqref{eq:Ito} with constant noise $\s (x,t) = s$ admits proper W-symmetries if and only if the drift term $f(x,t)$ is either independent of $x$ (case A) or depending linearly on $x$ (case B). In case $A$ we have an infinite set of W-symmetries for any value of $r \not= 0$, in correspondence to an arbitrary function of the variable
$$ \zeta \ := \ x \, - \, s \, w \, - \, H (t) $$ and identified by
$$ \vphi (x,t;w) \ = \ P [ x \, - \, s \, w \, - \, H (t) ] \ + \ r \ \[ x \, - \, H (t) \] \ , $$ while in case $B$, with $f(x,t)$ as in \eqref{eq:fB}, we have one W-symmetry for each arbitrary value of $r \not= 0$, given by \eqref{eq:Y} and identified by
\beq \vphi (x,t;w) \ = \ \b \ e^{K(t)} \ + \ r \, x \ - \  r \, e^{K(t)} \ \int_0^t e^{- K (\tau) } \ h(\tau) \ d \tau \ , \eeq with $\b$ an arbitrary constant.}

\medskip\noindent
{\bf Proposition 2.} (Standard symmetries) {\it An Ito equation \eqref{eq:Ito} with constant noise $\s (x,t) = s$ admits standard symmetries if and only if the drift term $f(x,t)$ falls in one of the three cases A,B,C given in Table \ref{tab:1}; the symmetries are then given, up to a multiplicative constant, by \eqref{eq:Y} with $\vphi (x,t;w)$ as given in Table \ref{tab:1}. In case $A$ this depends on an arbitrary function of the variable $\zeta$, while in cases $B$ and $C$ there is -- for $f(x,t)$ of an allowed form -- only one symmetry.}

\medskip\noindent
{\bf Remark 9.} The structure of proper W-symmetries described in Proposition 1 corresponds to the one expected on the basis of Remark 2; that is, we have the superposition of a strict proper W-symmetry and of a general standard symmetry. The strict proper W-symmetries are identified by $\omega (x,t,w)$ (see Remark 2) and in the different cases are as follows:
\begin{itemize}
\item In case $A$ -- where we have $f(x,t) = H'(t)$ -- we get $$ \omega_A (x,t;w) = r \, \[ x \ - \ H(t) \] \ ; $$
    \item In case $B$ -- where we have $f(x,t) = h (t) + K'(t) x$ -- we get $$ \omega_B (x,t;w) \ = \ r \, \[ x \ - \  e^{K(t)} \ \int_0^t e^{- K (\tau) } \ h(\tau) \ d \tau \] \ ; $$
    \item In case $C$ we have no proper W-symmetry.
\end{itemize}


\noindent Note that in case $B$, there is a hidden arbitrary constant, corresponding to the integration constant in the integral. \EOR

\medskip\noindent
{\bf Remark 10.} In the case of \emph{autonomous} Ito equation -- which in the present context means just $f(x,t) = f(x)$, as the noise term is not only autonomous but just constant -- the situation gets slightly simplified. This is summarized, for greater clarity and later reference, in Table \ref{tab:2} for what concerns \emph{standard} symmetries.

As noted in Remark 2 (see also Remark 9), the most general W-symmetry is a composition of two independent parts: the most general standard symmetry and a strict proper W-symmetry, identified by $\omega (x,t;w)$. Thus we only have to identify the latter, which is obtained by specifying the formulas given in  Remark 9 above. We get
\begin{itemize}
\item In case $A$, where now $f(x) = c$, we have $$ \omega_A (x,t;w) \ = \ r \, \[ x \ - \ c \, t \] \ ; $$
\item In case $B$, where now $f(x) = c_0 + c_1 x$, we have $$ \omega_B (x,t;w) \ = \  r \, \[ x \ + \  (c_0/c_1) \( 1 \ + \ \gamma \, e^{c_1 t} \) \] \ ; $$
\item In case $C$ we have of course no proper W-symmetry.
\end{itemize}


\noindent Note that in case $B$ there is an arbitrary constant $\gamma$, corresponding to the constant of integration for the formulas given in Remark 9.  \EOR
\bigskip

We will note here that in the case of autonomous equation and $s=1$ the drift coefficients in the three cases are respectively\footnote{Here we use $\Psi$ instead of $f$ for the drift coefficients and $\Phi$ instead of $\varphi$ for the symmetry ones to stress they correspond to special cases, and for later reference in Sect.\ref{sec:nonconst}.}
\begin{eqnarray}
\Psi_A (x)  &=& c \ , \nonumber \\
\Psi_B (x)  &=& c_0 \ + \ c_1 \ x \ , \label{eq:Psicases} \\
\Psi_C (x)  &=& c_0 \ + \ c_1 \ \exp[ \b x] \ ; \nonumber \end{eqnarray}
correspondingly the symmetry coefficients are
\begin{eqnarray}
\Phi_A (x,t;w) &=& P (x - w - c t) \ , \nonumber \\
\Phi_B (x,t;w) &=& \exp[ c_1 t] \ , \label{eq:Phicases} \\
\Phi_C (x,t;w) &=& \exp [\b (x - w - c_0 t)] \ . \nonumber \end{eqnarray}

\subsection{Discussion}

As mentioned above, case A is of little interest, in that in this case the equation is directly integrated.

In case B, our discussion shows that the only standard symmetries we obtain are deterministic ones; thus our result in this case just reproduces the one obtained by the Kozlov classification \cite{Koz2}.

On the other hand, in case C the only standard symmetries we obtain are \emph{random} ones, and these were not considered in Kozlov classification.

Similarly, we obtain proper W-symmetries -- {and hence also strict proper W-symmetries} -- only in the trivial case A and in case B; as these were not considered as well in Kozlov classification \cite{Koz2}, our result for case B is in fact new.

In the next Section \ref{sec:nonconst} we will show how our discussion -- and thus in particular these genuinely new results (compared with Kozlov classification) -- extends beyond the framework of equations with constant noise.

\begin{table}[h]
\centering
\begin{tabular}{||c||c|c||}
\hline
case & $f(x,t)$ & $\varphi(t,x; w)$\\[3pt]
\hline
\hline
A & $H'(t)$ & $P ( x - s w - H(t))  $
\\[3pt]
\hline
B & $a(t) + B'(t) x$ &
$ \exp [ B(t) ]$
\\[3pt]
\hline
C & $ A'(t) \, + \, b(t) \, \exp [\b x]$ &
$\exp[ \b (x  - s w - A(t))] $ \\[3pt]
\hline
\end{tabular}
\caption{The standard ($r=0$) symmetries when $\sigma(x)=s$, identified by $\vphi$ through eq. \eqref{eq:Ysimp}. In case A, $P$ is an arbitrary smooth function of its argument; in case C, $\b$ is a real constant. For the case with $r \not= 0$, see Remark 9.} \label{tab:1}
\end{table}

\begin{table}[h]
\centering
\begin{tabular}{||c||c|c||}
\hline
case & $f(x)$ & $\varphi(t,x; w)$\\[3pt]
\hline
\hline
A & $c$ & $P ( x - s w - c t)  $
\\[3pt]
\hline
B & $c_0 + c_1 x$ &
$ \exp [ c_1 t ]$
\\[3pt]
\hline
C & $ c_0 \, + \, c_1 \, \exp [\b x]$ &
$\exp[ \b (x  - s w - c_0 t)] $ \\[3pt]
\hline
\end{tabular}
\caption{The standard ($r=0$) symmetries when $\sigma(x)=s$ and for \emph{autonomous} Ito equations, identified by $\vphi$ through eq. \eqref{eq:Ysimp}. In case A, $P$ is an arbitrary smooth function of its argument; in case C, $\b$ is a real constant.  For the case with $r \not= 0$, see Remark 10.} \label{tab:2}
\end{table}

\section{Simple noise, autonomous Ito equations}
\label{sec:nonconst}

We will now consider the case of general noise coefficient. It will turn out, not surprisingly, that our results are more explicit in the case of \emph{autonomous equations} and even more for \emph{simple noise}; recall in this respect that, as mentioned in the Introduction, one can always reduce an Ito equation admitting a (deterministic) symmetry to the autonomous form \cite{Koz2}.

As these cases are also relevant in applications, we will start by considering them, i.e. set
\beq f(x,t) \ = \ f (x) \ , \ \ \ \s (x,t) \ = \ s \ x^k \ , \eeq and bound ourselves to standard symmetries; only in a second time we pass to consider the most general situation. Recall we will always assume $\s (x,t) \not= s$, i.e. $k \not= 0$ (as we discussed this case in the previous Section), and $\s (x,t) \not= s x$, i.e. $k \not= 1$, as this case was discussed previously in the literature \cite{GSclass}. Needless to say, we will always assume $s \not= 0$, or the equation would be deterministic rather than stochastic.

We will follow the approach described in Section \ref{sec:standard}. We will thus define a new variable $y$ through \eqref{eq:Ydef}, which in the present case of simple noise reads
\beql{eq:DCC} y \ = \ g(x) \ = \ \int \frac{1}{\s (x)} \ d x \ = \ \frac{x^{1-k}}{s \, (1-k)} \ ; \eeq
note that in the last equality we have taken advantage of the explicit form of $\s (x)$. We note, for later reference, that the inverse change of variables is just
\beql{eq:ICC} x \ = \ [ (1-k) \, y]^{1/(1-k)} \ := \ \xi (y) \ . \eeq

As discussed in Section \ref{sec:standard}, in this way our equation is transformed into
\beql{eq:INF} d y \ = \ F(y) \, dt \ + \ d w \ , \eeq where the function $F(y)$ is given explicitly by \eqref{eq:Fgenyt}, which in the autonomous case reduces to \eqref{eq:Fgenaut}.
In our case $\pa \s / \pa x = s k x^{k-1}$; using \eqref{eq:ICC} this reads
$$ \(\frac{\pa \s}{\pa x} \) \ = \ s \, k \, \[ (1 -k) \, y \]^{-1} \ = \ \( \frac{s \, k}{1-k} \) \ \frac{1}{y} \ . $$ Thus we get
\beql{eq:H} F (y) \ = \ \frac{f [\xi (y)]}{s \ [\xi(y)]^k} \ - \ \frac12 \ \( \frac{s \, k}{1-k} \) \, \frac{1}{y} \ . \eeq

Now, the relevant point is that \eqref{eq:INF} is an Ito equation with \emph{constant noise}, i.e. of the type which was considered in Section \ref{sec:const}. Thus we know that it admits symmetries if and only if the drift coefficient $F(y)$ is in one of the three allowed forms identified in our discussion there, and actually with $s=1$, see Table \ref{tab:2}.

\subsection{Equations admitting symmetries}
\label{sec:admitting}

Our task is then to understand which form of $f(x)$ leads to an $F(y)$ of the allowed form. We write these allowed forms (with $s=1$) as $\Psi_\a (y)$, $\a = A,B,C$.

Now we also note that our computations are simpler if we work in the $x$ variable, i.e. if instead of using \eqref{eq:ICC} to express $x$ in the terms of the new variable $y$, we use the direct change of variables \eqref{eq:DCC} to write $y = g (x)$. In this way the equations to be satisfied are just
\beq \frac{f(x)}{\s(x)} \ - \ \frac12 \ \s' (x) \ = \ \Psi_a [g(x)] \ , \eeq
which of course means
\beql{eq:3NF} f(x) \ = \ \[ \Psi_a [ g(x)] \ + \ \frac12 \ \s' (x) \] \ \s (x) \ . \eeq

\subsubsection{Case A}

The first case is also the simplest one. Now $\Psi (y) = \Psi_A (y) = c_0 $, and eq.\eqref{eq:3NF} yields
\beq f(x) \ = \ \[ c_0 \ + \ \frac12 \ \s' (x) \] \ \s (x) \ . \eeq In the simple noise case, i.e. using \eqref{eq:DCC}, this means
\beq f(x) \ = \ c_0 \, s \, x^k \ + \ \frac{s^2 \, k}{2} \, x^{2k-1} \ . \eeq

\subsubsection{Case B}

In the second case we have $\Psi (y) = \Psi_B (y) = c_0 + c_1 y$ and we have
\beq f(x) \ = \ \[ c_0 \ + \ c_1 \ g(x) \ + \ \frac12 \, \s' (x) \] \ \s (x) \ . \eeq
In the simple noise case this reads (recall we set $k \not= 1$)
\begin{eqnarray} f(x) &=& \[ c_0 \ + \ \frac{c_1}{s \, (1-k)} \ x^{1-k} \ + \ \frac{s \, k}{2} \, x^{k-1} \] \ s \, x^k \nonumber \\
&=& c_0 \, s \, x^k \ + \ \frac{c_1}{(1-k)} \, x \ + \ \frac{s^2 \, k}{2} \, x^{2k-1}
 \ . \end{eqnarray}

\subsubsection{Case C}

In the third case we have $\Psi (y) = \Psi_C (y) = c_0 + c_1 \exp [y]$ and we have
\beq f(x) \ = \ \[ c_0 \ + \ c_1 \ \exp[ g (x)] \ + \ \frac12 \, \s '(x) \] \ \s (x) \ . \eeq
In the simple noise case, and recalling once again $k \not= 1$, this reads
\beq f(x) \ = \ c_0 \, s \, x^k \ + \ \frac{s^2 \, k}{2} \, x^{2 k - 1}
\ + \ c_1 \, s \, x^k \, \exp \[ \frac{A}{s \, (1-k)} \ x^{1 - k} \] \ . \eeq

\subsection{Admitted symmetries}
\label{sec:admitted}

We have thus classified all the autonomous scalar Ito equations admitting symmetries. In order to obtain the explicit form of the admitted symmetries we have two different ways to proceed. On the one hand, we could use the fact that symmetries of Ito equations are conserved under a change of variables, as shown by Lunini \cite{GL1,GL2}; or we can directly solve the determining equations \eqref{eq:deteq1}, \eqref{eq:deteq2} for each of these.

\medskip\noindent
{\bf Remark 11.} The result of Lunini was obtained before the introduction of W-symmetries, so it is not clear that it would apply in this more general framework. Actually, it turns out this is \emph{not} the case, as will appear from our direct computations in Appendix \ref{app:noW}. The direct proof of conservation of standard (deterministic or random) symmetries provided in Appendix \ref{app:symmcons} for the simple scalar case will also help clarifying this point. \EOR
\bigskip

In the present case of simple noise, once we know both $f$ and $\s$, determining the symmetry vector fields turns out to be a simple task. We will thus proceed in this way.

Actually, a part of the computation can be made without reference to the different cases: in fact, the equation \eqref{eq:deteq2} only depends on $\s$, and not on $f$.

For our present assumption $\s (x,t) = s x^k$, and with $r=0$ (i.e. looking for standard symmetries) the equation \eqref{eq:deteq2} reads
\beq \vphi_w \ + \ s \, x^k \, \vphi_x \ - \ s \, k \, x^{k-1} \, \vphi \ = \ 0 \ . \eeq
This is promptly solved by the method of characteristics, yielding
\beql{eq:phiNSN} \vphi (x,t;w) \ = \ x^k \ \psi (z,t) \ , \eeq
where we have defined
\beql{eq:zNSN} z \ := \ w \ + \ \frac{x^{1-k}}{s \, (k-1)} \ . \eeq

Now the equation \eqref{eq:deteq2} is identically satisfied, while \eqref{eq:deteq1} becomes an equation for $\psi$, depending on the form of $f(x,t)$.

\subsubsection{Case A}

In case A we have
$$ f(x,t) \ = \ c \, x^k \ + \ \frac12 \,  s^2 \, k \, x^(2 k - 1) \ ; $$
the equation \eqref{eq:deteq1} is then
\beq \frac{x^k}{s} \ \( s \, \psi_t \ - \ c \, \psi_z \) \ = \ 0 \ . \eeq
The general solution to this reads
\beq \psi (z,t) \ = \ \eta (u) \ , \ \ \ \ u := c\, t \ + \ s \, z \ , \eeq
and in conclusion we have symmetries depending on an arbitrary function, i.e.
\beq \vphi_A (x,t;w) \ = \ x^k \ \eta (u) \ . \eeq

\subsubsection{Case B}

In case B, we have
$$ f(x,t) \ = \ c_0 \, s \, x^k \ + \ \frac{c_1}{1 - k} \,  x + \frac12 \, s^2 \, k \, x^{2 k - 1} \ ; $$
the equation \eqref{eq:deteq1} is then of the form
$$ \frac{c_1}{(k-1) \, s} \ x \ \psi_z \ + \ x^k \ \Theta [ \psi (x,t) ] \ = \ 0 \ , $$ with $\Theta$ a linear operator which we do not write down explicitly. This enforces
\beq \psi (z,t) \ = \ \eta (t) \ ; \eeq
now the equation \eqref{eq:deteq1} is reduced to
\beq x^k \ \[ \eta' (t) \ - \ c_1 \ \eta (t) \] \ , \eeq
which yields immediately
$$ \eta (t) \ = \ \exp [ c_1 t ] $$ and hence
\beq \vphi_B (x,t;w) \ = \ x^k \ e^{c_1 t} \ . \eeq

\subsubsection{Case C}

Finally, in case C we have
$$ f(x,t) \ = \ c_0 \, x^k \ + \ c_1 \, x^k \exp[\beta \, x^{1 - k}] \ + \ \frac12 \, s^2 \, k \, x^{2 k - 1} \ . $$ The equation \eqref{eq:deteq1} reads now
\beq \frac{x^k}{s} \ \[ s \, \psi_t \ - \ c_0 \, \psi_z \ + \ \exp[\b \, x^{1 - k}] \ \( \b \, c_1 \, s \, (k-1) \, \psi \ - \ c_1 \, \psi_z \) \] \ = \ 0 \ . \eeq
Requiring the vanishing of the part without the exponential yields
\beq \psi (z,t) \ = \ \eta (u) \ , \ \ \ \ u \ := \ t + (s/c_0) z \ ; \eeq
plugging this into the full equation we get (omitting the exponential factor)
\beq \b \, c_1 \, s \, (k-1) \ \eta \ - \ \frac{c_1 \, s}{c_0} \ \eta' \ = \ 0 \ , \eeq
and hence
\beq \eta (u) \ = \ K \ \exp [ \b \, c_0 \, (k-1) \, u ] \eeq
with $K$ an arbitrary constant. In conclusion,
\beq \vphi_C (x,t;w) \ = \ K \, x^k \ \exp [ \b \, c_0 \, (k-1) \, u ] \ . \eeq

\subsection{Admitted symmetries, the alternative approach}
\label{sec:admittedalter}

As stated earlier on, standard symmetries of an Ito equation are preserved under a change of variables. We can use this fact to determine in a different way (but obtaining the same results) the symmetries for the symmetric equations identified in Section \ref{sec:admitting}.

We should recall that the equations identified by the $f(x)$ listed in Tables \ref{tab:3} and \ref{tab:4} are mapped into those listed in Table \ref{tab:2} by the map $x \to y = g(x)$, and that in the presently considered case of simple noise $\s (x) = s x^k$, we have $g(x) \ = \ x^{1-k}/( (1-k) s)$.
Thus the symmetries correspond to $\Phi_k [g(x),t;w]$ with $\Phi_k$ ($k=A,B,C$) the functions listed in \eqref{eq:Phicases}. Note that the symmetry vector field is $Y = \Phi_k (y,t;w) \pa_y$; as we want to express this in terms of $x$, hence of $\pa_x$, we should recall that $\pa_x = (\pa g/ \pa x) \pa_y = (1/\s (x)) \pa_y$, i.e. that $\pa_y = \s (x) \pa_x = x^k \pa_x$ (the inessential constant $s$ has been dropped).

One readily checks that the structure of the symmetries we have determined in the previous section \ref{sec:admitted}, see also Tables \ref{tab:3} and \ref{tab:4}, is exactly this.

\subsection{Summary of results for simple noise}

We can summarize the results obtained for general noise in Table \ref{tab:3}; in this case we can only give results in terms of the generic function $g(x)$, to be computed through the integral in \eqref{eq:g} (see \eqref{eq:DCC} as well), so we have an implicit form classification.

In the case of simple noise \eqref{eq:sn} we can obtain the classification in explicit form; this is provided in Table \ref{tab:4}.

\medskip\noindent
{\bf Remark 12.} In this Section we only discussed standard symmetries, and not W-symmetries. It turns out that for (non-constant) simple noise, and under our assumption $k \not= 1$, there is no proper W-symmetry. This assertion is proven in detail in Appendix \ref{app:noW}. \EOR


\begin{table}[h]
\centering
\begin{tabular}{||c||c|c||}
\hline
case & $f(x)/\s (x)$ & $\varphi(t,x; w)/\s (x)$ \\[3pt]
\hline
\hline
A & $c  + (1/2) \s_x (x) $ & $P [ g(x) - c t -  w]  $
\\[3pt]
\hline
B & $c_0 + c_1 g(x) + (1/2) \s_x (x)  $ &
$ \exp [ c_1 t ]$
\\[3pt]
\hline
C & $ c_0  +  c_1 \exp [\b g(x)] + (1/2) \s_x (x)  $ &
$\exp[ \b (g(x) - w - c_0 t)] $ \\[3pt]
\hline
\end{tabular}
\caption{The standard ($r=0$) symmetries for general autonomous noise term $\sigma(x)$, identified by $\vphi$ through eq. \eqref{eq:Y}. In case A, $P$ is an arbitrary smooth function of its argument; in case C, $\b$ is a real constant. The function $g(x)$ is given by \eqref{eq:g}. We tabulate $f(x)/\s (x)$ and $\varphi (x,t;w) / \s (x)$ for typographical clarity.} \label{tab:3}
\end{table}

\begin{table}[h]
\centering
\begin{tabular}{||c||c|c||}
\hline
case & $f(x)$ & $\varphi(t,x; w)$\\[3pt]
\hline
\hline
A & $c x^k + (1/2) s^2 k x^{2 k -1} $ & $x^k P [ x^{1-k}/(1-k) - s w - c t]  $
\\[3pt]
\hline
B & $c_0 x^k + c_1 x + (1/2) s^2 k x^{2 k - 1} $ &
$ x^k \, \exp [ (1-k) c_1 t ]$
\\[3pt]
\hline
C & $ c_0 x^k +  c_1 x^k \exp [\b x^{1-k}] + (1/2) s^2 k x^{2 k - 1} $ &
$x^k \, \exp[ \b (x^{1-k} - (1-k) (s w + c_0 t))] $ \\[3pt]
\hline
\end{tabular}
\caption{The standard ($r=0$) symmetries for simple $\sigma(x) = s x^k$ (with $s \not= 0$ and $k \not= 0 $, $k \not= 1$), identified by $\vphi$ through eq. \eqref{eq:Y}. In case A, $P$ is an arbitrary smooth function of its argument; in case C, $\b$ is a real constant.} \label{tab:4}
\end{table}


\section{The general situation: non-autonomous equations, non-simple noise}
\label{sec:nonconstgen}

The approach pursued in the previous Section \ref{sec:nonconst} for simple noise and autonomous equations can also be applied in the general case of non-autonomous equations and general noise (always excluding the cases $\s (x,t) = s$ and $\s (x,t) = s x$).

In such general case, obviously, we will not be able to write explicitly the direct and inverse changes of coordinates \eqref{eq:DCC} and \eqref{eq:ICC}. That is, albeit we will be able to classify symmetric equations and their standard symmetries, this classification will be only implicit, and for each noise term $\s (x,t)$ one will have to perform the integral appearing in \eqref{eq:DCC} and attempt to invert this in order to get the equivalent of \eqref{eq:ICC}.

We stress that for $\s$ depending on $t$, we have to use the full form \eqref{eq:Fgenyt} of the transformed drift coefficient, rather than the simplified form \eqref{eq:Fgenaut}. Moreover, as we started from a non-autonomous equation, the transformed (constant noise) equation will in general be also $t$-dependent; in other words, we have to use the $\Psi_a (y,t)$ identified by Table \ref{tab:1} instead of the $\Psi_a (y)$ identified by the simplified Table \ref{tab:2}.

The equation to be solved in order to identify the $f(x,t)$ giving raise to $F(y) = \Psi_a (y)$ under the change of coordinates \eqref{eq:DCC} are then
\beql{eq:fgen0} f(x,t) \ = \ \[ \Psi_a [g(x,t),t] \ + \ \frac12 \ \s_x (x,t) \ - \ \int \frac{\s_t (x,t)}{\s^2 (x,t)} \ dx \] \ \s (x,t) \ . \eeq
We note that this can be written in a more compact way; in fact, the last term within the square brackets is also written as $\pa_t g(x,t)$. Thus \eqref{eq:fgen0} reads also
\beql{eq:fgen} f(x,t) \ = \ \Psi_a [g(x,t),t] \ \s (x,t) \ + \ \frac14 \, [\pa_x \s^2 (x,t) ] \ + \ \s (x,t) \, \ \pa_t [ g (x,t) ] \ . \eeq

This formula, together with the explicit form of the $\Psi_a (y,t)$ provided in Table \ref{tab:1} and \eqref{eq:g}, contains the whole classification. As already remarked, for a general functional form of $\s (x,t)$ we cannot provide a more explicit representation of the allowed $f(x,t)$.

We thus summarize our discussion in the following Proposition; unfortunately, it gives the result in a rather implicit way. Some explicit examples, which illustrate how these Propositions can be used in practice once one gives a concrete noise term $\s = \s (x,t)$, are given in Appendix \ref{app:exagen}.

\medskip\noindent
{\bf Proposition 3.} {\it A general Ito equation \eqref{eq:Ito} admits a standard symmetry if and only if the drift and noise terms $f(x,t)$ and $\s (x,t)$ satisfy the relations embodied in \eqref{eq:fgen} and \eqref{eq:g} for $\Psi_a (y,t)$ one of the functions described in Table \ref{tab:1}. In this case the symmetries are provided by the functions $\Phi_a (y,t;w)$ given in Table \ref{tab:1}.  }
\bigskip

The situation is slightly simpler -- but still defying an explicit classification -- in the case of autonomous equations with a general (not simple) noise. Now the relevant functions $\Psi_a (y)$ are those of Table \ref{tab:1}, and the equations to be satisfied are
\beql{eq:fgenaut} f(x) \ = \ \Psi_a [g(x)] \ \s (x) \ + \ \frac14 \ \pa_x \s^2 (x)   \ , \eeq
\beql{eq:ggenaut} g(x) \ = \ \int \frac{1}{\s (x)} \ d x \ . \eeq

\medskip\noindent
{\bf Proposition 4.} {\it A general \emph{autonomous} Ito equation \beql{eq:Itoaut} dx \ = \ f(x) \, dt \ + \ \s (x) \, d w \eeq
admits a standard symmetry if and only if the drift and noise terms $f(x)$ and $\s (x)$ satisfy the relations embodied in \eqref{eq:fgenaut} and \eqref{eq:ggenaut} for $\Psi_a (y)$ one of the functions described in Table \ref{tab:2}. In this case the symmetries are provided by the functions $\Phi_a (y,t;w)$ given in Table \ref{tab:2}. }

\medskip\noindent
{\bf Remark 13.} Proposition 3 and Proposition 4 describe the situation for general (i.e. non.autonomous) and autonomous Ito equations respectively in a rather implicit fashion. In Table \ref{tab:5}, where we use the shorthand notations \eqref{eq:g}, \eqref{eq:varrho}, a somewhat more direct description is provided. Note this Table describes the general situation (non autonomous Ito equations) but also applies to the case of autonomous equations: in this case one has to specialize to $f$ and $\s$ independent of $t$; needless to say, now $g(x,t)$ and $\varrho (x,t)$ are actually functions of $x$ alone as well. \EOR

\medskip\noindent
{\bf Remark 14.} It should be noted that the notation in terms of indefinite integrals, as in \eqref{eq:g}, \eqref{eq:varrho}, is potentially confusing in some situation as we don't keep track of the integration constants. A more precise notation would be as
\beql{eq:M1Tgrho} g(x,t) \ := \ \int_{x_0}^x \frac{1}{\s (x',t)} \, d x' \ , \ \ \
\varrho (x,t) \ := \ \int_{x_0}^x \frac{\s_t (x',t)}{\s^2 (x',t)} \, d x' \ , \eeq
In discussing eq.\eqref{eq:ICCC} below we will have to be a bit more careful about this point. \EOR
\bigskip

\begin{table}[h]
\centering
\begin{tabular}{||c||c|c||}
\hline
case & $f(x,t)/\s (x,t)$ & $\varphi(x,t; w)/\s (x,t)$\\[3pt]
\hline
\hline
A & $H'(t)+ \frac12  \sigma_x (x,t)+  \varrho (x,t) $ & $P[g(x,t)- w-H(t)]$
\\[3pt]
\hline
B & $a(t)+B'(t)g(x,t)+ \frac12  \sigma_x (x,t)+ \varrho (x,t) $ & $\exp[B(t)]$
\\[3pt]
\hline
C &  $A'(t)+ b(t)\exp[\b g(x,t)]+ \frac12  \sigma_x (x,t)+ \varrho (x,t) $ & $\exp[\b(g(x,t) - w - A(t))]$\\[3pt]
\hline
\end{tabular}
\caption{The standard ($r=0$) symmetries for general (i.e. non autonomous) Ito equation and arbitrary noise $\sigma(x,t)$. This table uses the shorthand notation \eqref{eq:M1Tgrho}, and we tabulate $f/\s$ and $\vphi/\s$,  for typographical convenience.}
\label{tab:5}
\end{table}

\section{Integration of symmetric equations}
\label{sec:integration}

It is rather clear than once we have integrated the symmetric equations with constant noise identified in Section \ref{sec:const}, a simple use of the changes of coordinates \eqref{eq:DCC}, \eqref{eq:ICC} allows also to integrate the symmetric equation with non-constant noise identified in Section \ref{sec:nonconst}.

We will thus just focus on the case with constant noise. The solutions for the integrable equations with non-constant noise can be obtained from these using the transformations taking those equations into their standard form with constant (unit) noise, as discussed at length in Section \ref{sec:standard}.


In order to integrate the symmetric scalar Ito equations, we will use the Kozlov approach \cite{\kozref}, which amounts to a symmetry-adapted change of coordinates (see Appendix \ref{app:Koz} for details). This will of course be different for each one of the three cases we have identified in Section \ref{sec:const}.

\subsection{Case A}

In case A we have equations of the form
\beq d x \ = \ a(t) \, dt \ + \ s \, d w \ ; \eeq
these are immediately integrated with no need to use symmetries, providing
\beq x(t) \ = \ x(t_0) \ + \ \int_{t_0}^t  a (\tau) \, d \tau \ + \ s \ [ w(t) \, - \, w (t_0 ) ] \ . \eeq

\subsection{Case B}
\label{sec:intB}

In case B we have equations of the form
\beq d x \ = \ [ a(t) \, + \, b(t) \, x ] \, dt \ + \ s \, d w \ ; \eeq
these admit the symmetry
\beq Y \ = \ \exp[ B (t)] \pa_x \ , \ \ \ \ B(t) \ = \ \int b(t) \, d t \ , \eeq
corresponding to $\vphi (x,t;w) = \exp[ B(t)]$.

We change variable passing to
\beql{eq:CVB} y \ = \ \int \frac{1}{\vphi} \ d x \ = \ e^{- B(t) } \ x \ . \eeq
This satisfies (note that as the change of variable is linear in $x$ the Ito Laplacian gives a null contribution)
\begin{eqnarray*}
dy &=& e^{- B(t)} d x \ - \ b (t) \ e^{- B(t)} \ x \ d t \\
&=& e^{- B (t)} \ \[ \( a(t) \, + \, b(t) \, x \) \, dt \ + \ s \, d w \] \ - \ b (t) \ e^{- B(t)} \ x \ d t \\
&=&  a(t) \, e^{- B(t)} \ d t   \ + \ s \ d w \ ; \end{eqnarray*}
this is promptly integrated to give
\beq y(t) \ = \ y(t_0) \ + \ \int_{t_0}^t a(\tau ) \, e^{- B(\tau )} \, d \tau \ + \ s \ \[ w(t) \, - \, w (t_0 ) \] \ . \eeq
We then invert the change of variable \eqref{eq:CVB}, which gives
\beq x \ = \ e^{B(t)} \ y \eeq
(again we have no contribution from the Ito Laplacian, as we have to deal with a linear transformation), so that
\beq x(t) \ = \ x (t_0) \ + \ e^{B(t)} \ \[ \int_{t_0}^t a(\tau ) \, e^{- B(\tau )} \, d \tau \ + \ s \ \[ w(t) \, - \, w (t_0 ) \] \] \ . \eeq

\subsection{Case C}
\label{sec:intC}

In case C we have equations of the form
\beq d x \ = \ \[ a(t) \, + \, b(t) \, e^{\b x} \] \, d t \ + \ s \, d w \ ; \eeq
these admit the symmetry
\beq Y \ = \ \exp \[ \b \, \( x - s w - A(t) \) \] \ \pa_x \ . \eeq
The Kozlov change of variables is now
\beq y \ = \ \int e^{- \b (x - s w - A(t) )} \ d x \ = \ - \, \frac{1}{\b} \ e^{- \b (x - s w - A(t) )} \ , \eeq
with inverse
\beql{eq:ICCC} x \ = \ - \, \frac{1}{\b} \ \log (\b \, y) \ + \ s \, w \ + \ A(t) \ . \eeq

\medskip\noindent
{\bf Remark 15.} This formula is seemingly singular due to the presence of the logarithm term. As anticipated in Remark 14 above, this is due to our compact notation with indefinite integrals. In fact, if we use the definitions given in \eqref{eq:M1Tgrho},
then we obtain
$$ y-y_0 \ = \ \frac{1}{\beta} \ e^{\beta(sw+A(t))} \ \left(e^{-\beta x_0}-e^{-\beta x}\right) $$ and hence
$$ e^{-\beta x} \ = \ e^{-\beta x_0} \ - \ \beta \ e^{-\beta(sw+A(t))}(y-y_0) \ . $$ Thus, in the end,
\beq x \ = \ - \ \frac{1}{\beta } \ \log\left[e^{-\beta \, (x_0 - s w - A(t))} \ + \ \beta \, (y_0 - y) \right] \ + \ s \, w \ + \ A(t) \ . \eeq
Since possible sources of confusion are clarified, we will revert to the compact notation. \EOR
\bigskip

Now we should compute
$$ \Delta (y) \ = \ \[ - \b \, s^2 \ + \  2 \, \b \, s^2 \ - \ \b \, s^2 \] \ e^{- \b (x - s w - A(t) )} \ = \ 0 \ . $$
Thus we have
\begin{eqnarray*}
dy &=& \frac{\pa y}{\pa x} \ d x \ + \ \frac{\pa y}{\pa t} \ d t \ + \ \frac{\pa y}{\pa w} \ d w \\
&=& e^{- \b (x - s w - A(t) )} \ \[ d x \ - \ s dw \ - \ a(t) \, d t \] \\
&=& e^{- \b (x - s w - A(t) )} \ \[ \( a(t) + b(t) \, e^{\b x} \) \, dt \ + \ s \, dw \ - \ s \, dw \ - a(t) \, dt \] \\
&=& e^{\b (s w + A(t) )} b(t) \ d t \ . \end{eqnarray*}

This shows that -- as it generally happens -- the transformed equation is \emph{not} of Ito type. We still have
\beq y(t) \ = \ y(t_0) \ + \ \int_{t_0}^t e^{\b [s w(\tau)  + A(\tau) ]} b(\tau ) \ d \tau \ , \eeq
and $x(t)$ is recovered by \eqref{eq:ICCC}.

\medskip\noindent
{\bf Remark 16.} As discussed in \cite{GL2} (see Theorem 5 therein) one can know \emph{a priori} if the Kozlov reduction of an Ito equation \eqref{eq:Ito} under a random standard symmetry \eqref{eq:Ysimp} will produce an Ito type equation. In fact, defining the function $\ga (x,t;w)$ as $\ga := \pa_w (1/\vphi)$, this is the case if and only if the equation
$$ \s \, \ga_t \ + \ \s_t \, \ga \ = \ f \, \ga_w \ + \ \frac12 \ \( \s \, \ga_{ww} \ + \ \s^2 \, \ga_{xw} \) $$
is satisfied.

In the present case, with the $f(x,t)$ and $\vphi (x,t;w)$ as in Table \ref{tab:1} (case $C$), and setting $s=1$ (obviously nothing changes with a general $s \not= 0$), a direct computation shows that this equation is equivalent to
$$ \b^2 \ b (t) \ \exp \[ w \ + \ A (t) \] \ = \ 0 \ , $$
i.e. is never satisfied, except in the trivial cases $\b = 0$ or $b(t) = 0$; note that in both these cases we are actually back to case $A$. \EOR

\section{Conclusions}

We have thus completed our task. That is, we provided a classification of symmetries for scalar Ito stochastic differential equations. This is based on the one hand on the classification for such equations with constant noise $\s (x,t) = s$, obtained in Section \ref{sec:const}; and on the other on the (well known) possibility to reduce any non-singular Ito equation to this standard case by means of the transformations discussed in Section \ref{sec:standard}.

This kind of classification was considered previously by Kozlov \cite{Koz2}; his results are on the one hand more extended, as he considers also transformations acting on time; and on the other hand less extended, as at the time of his work certain types of symmetries, i.e. random standard symmetries and -- albeit in the end we found they are quite rare -- W-symmetries, had not yet been introduced in the literature. Thus our work does add to Kozlov's work the classification of equations possessing these types of symmetries.

All in all, it is quite clear from our Tables that -- despite considering this wider set of admitted transformations -- symmetric (scalar) Ito equations are a very small subset in the class of (scalar) Ito equations. This is not surprising: even in the deterministic setting symmetry is a highly non-generic property (albeit physicists may have a different feeling when considering equations originating from fundamental Physics, due to the inherent isotropy and homogeneity of space and time-translation invariance), and in the stochastic case symmetries must preserve both the drift -- i.e. roughly speaking the deterministic part of the equation -- and the noise coefficients.

It should be stressed that, as already stated in the Introduction, the list of symmetric equations is rather limited, but not exceedingly poor; moreover, even in the simpler case of constant noise and standard symmetries (see Table \ref{tab:2}), when one takes into account standard random symmetries as well, it includes equations with exponential drift -- a feature which is rather surprising if compared to previously known results.

On the other hand, we have now identified symmetric equations, and discussed in Section \ref{sec:integration} how the symmetries are used to integrate them. This provides a ``ready to use'' catalogue of problems which can be exactly integrated with no further effort; needless to say, they and their integration can also be used as a starting point for the perturbation study of ``nearby'' stochastic problems.

We also note that, albeit we focused mainly on standard symmetries -- as in this case the Kozlov map is guaranteed to transform the original equation into a new Ito equation, see Remark 1 -- we also partially discussed the case of W-symmetries, albeit to find out they are present only in the case of constant noise; see Remarks 2 and 9 above. We stress that the last statement should be understood within the limits of this study: recall we have excluded from consideration of simple noise the case of $\s(x,t) = s x$, as this was studied previously in the literature \cite{GSclass}; in this case, there are instead proper W-symmetries.

The obvious limitation of our study is that it concerns only \emph{scalar} equations. For an $n$-dimensional Ito equation, i.e. a system of $n$ coupled scalar Ito equations, the two determining equations \eqref{eq:deteq1}, \eqref{eq:deteq2}, are replaced by a system of $n + n^2$ similar equations (more precisely, $n$ second order PDEs and $n^2$ first order PDEs), so the difficulty of obtaining a classification increases quickly with the dimension of the system.

\addcontentsline{toc}{section}{Acknowledgements}

\subsection*{Acknowledgements}

We are very grateful to our two Referees for corrections, useful suggestions, and more generally for constructive criticism and rapid reports.
This work was developed in the course of a semester-long stay of MAR in Milano under the Program ``Salvador de Madariaga''. The support of Spain's Ministerio de Ciencia, Innovaci\'on y Universidades under grant PGC2018-094898-B-I00, as well as of Universidad Complutense de Madrid under grant G/6400100/3000 are also acknowledged. The work of GG is also supported by GNFM-INdAM.

\renewcommand{\theequation}{\thesection.\arabic{equation}}
\setcounter{equation}{0}
\appendix 


\section{The Kozlov transformation}
\label{app:Koz}

As mentioned in Section \ref{sec:geom}, once an Ito equation admits a symmetry  of the form \eqref{eq:Ito}, one can perform a change of variables mapping the equation into a new equation in which the drift and noise coefficients do not depend on $x$ -- in case the symmetry is a deterministic one, they only depend on $t$ -- and which can hence be explicitly (and immediately) integrated. This result is due to Kozlov \cite{Koz2} and is at the basis of the application of the symmetry approach to integrate stochastic equations; it is well known, but we discuss it here for the sake of completeness.

We stress that the converse is also true: if the equation can be integrated, being possible to bring it to an integrable form, then it must admit a standard symmetry.

That is, we want to prove that the Ito equation \eqref{eq:Ito} can be mapped into the target -- and immediately integrable -- form
\beql{eq:app:target} dy \ = \ F(t) \, d t \ + \ S(t) \, d w  \eeq
if and only if it admits a standard deterministic symmetry, i.e. a symmetry of the form \eqref{eq:Ysimp} where $\vphi_w = 0$.

Moreover, it can be mapped into the target non-Ito -- but also integrable -- form
\beql{eq:app:target:2} dy \ = \ F(t,w) \, d t \ + \ S(t,w) \, d w  \eeq
if and only if it admits a general standard symmetry \eqref{eq:Ysimp}.

\subsection{The Kozlov map}

We recall preliminarily that if the evolution of the random variable $x = x(t)$ is governed by the Ito equation
\beq d x \ = \ f(x,t) \, dt \ + \ \s (x,t) \, d w \label{eq:appA:Ito} \eeq and we operate a change of variable by
$$ y \ = \ \Phi (x,t;w) $$ (note that both $t$ and $w$ are unaffected by the change of variable) then the evolution of $y = y(t)$ is governed by
$$ dy \ = \  \frac{\pa \Phi}{\pa x } \ d x \ + \ \frac{\pa \Phi}{\pa t } \ d t \ + \ \frac{\pa \Phi}{\pa w } \ d w \ + \ \frac12 \ \De (\Phi) \ , $$ where as usual $\De$ is the Ito Laplacian.

We assume that the change of variables is non-singular (that is, $\Phi_x \not= 0$) and denote the inverse change of variables by
$$ x \ = \ \xi (y,t;w) \ ; $$ we have of course also $\xi_y \not= 0$.

Taking into account the equation for $dx$, we have
\begin{eqnarray*}
dy &=& \( \frac{\pa \Phi}{\pa x } \ f(x,t) \ + \ \frac{\pa \Phi}{\pa t } \ + \ \frac12 \De (\Phi) \) \ dt \ + \ \( \frac{\pa \Phi}{\pa x } \ \s (x,t) \ + \  \frac{\pa \Phi}{\pa w } \) \ d w \\
&:=& F(x,t;w) \ dt \ + \ S (x,t;w) \ dw \\
&=& F [ \xi (y,t;w),t;w] \ dt \ + \ S [ \xi (y,t;w) , t ;w] \ dw \\
&:=& \wt{F} (y,t;w) \ dt \ + \ \wt{S} (y,t;w) \ dw \ . \end{eqnarray*}

Now we note that, as $t$ and $w$ are not touched by our changes of variables,
$$ \frac{\pa}{\pa y} \ = \ \( \frac{\pa \xi}{\pa y} \) \ \frac{\pa}{\pa x} \ ; $$ in particular, this means that
\begin{eqnarray} & & \frac{\pa \wt{F}}{\pa y} \ = \ 0 \ \Longleftrightarrow \ \frac{\pa F}{\pa x} \ = \ 0 \ ; \nonumber \\
& & \\
& & \frac{\pa \wt{S}}{\pa y} \ = \ 0 \Longleftrightarrow \ \frac{\pa S}{\pa x} \ = \ 0 \ . \nonumber \end{eqnarray}

In other words, we can check the independence of $\wt{F}$ and $\wt{S}$ on $y$ by checking the independence of $F$ and $S$ on $x$. We note in passing that if $F_x=0$, $S_x=0$ -- and hence $\wt{F}_y = 0$, $\wt{S}_y = 0$, then $\wt{F} = F$, $\wt{S} = S$.

Now let us introduce the function
$$ \vphi (x,t;w) \ := \ \frac{1}{\pa \Phi / \pa x } \ ; $$ with this, we can also write
\beql{eq:app:kozmap} \Phi (x,t;w) \ = \ \int \frac{1}{\vphi(x,t;w) } \ dx \ . \eeq
We can rewrite $F (x,t;w)$ and $S(x,t;w)$ defined above in terms of $\vphi$. Using
\begin{eqnarray*}
\Phi_x &=& 1/\vphi \ , \ \
\Phi_t \ = \ - \ \int (\vphi_t / \vphi^2) \ dx \ , \ \
\Phi_w \ = \ - \ \int (\vphi_w / \vphi^2 ) \ d x \ ; \\
\Phi_{xx} &=& - \ \vphi_x / \vphi^2 \ , \ \
\Phi_{xw} \ = \ - \ \vphi_w/\vphi^2 \ , \ \
\Phi_{ww} \ = \ \int \[ (2 \vphi_w^2/\vphi^3) \ - \ (\vphi_{ww}/\vphi^2) \] \ d x \ , \end{eqnarray*}
we obtain promptly
\begin{eqnarray}
F &=& \frac{f}{\vphi} \ - \ \( \int \frac{\vphi_t}{\vphi^2} \ dx \) \nonumber \\
& & \ - \ \frac12 \ \[ \s^2 \, \frac{\vphi_x}{\vphi^2} \ + \ 2 \, \s \, \frac{\vphi_w}{\vphi^2} \ + \ \int \( \frac{\vphi_{ww}}{\vphi^2} \ - \ \frac{2 \, \vphi_w^2}{\vphi^3} \) \ dx \] \ , \label{eq:app:NF} \\
S &=& \frac{\s}{\vphi} \ - \ \int \frac{\vphi_w}{\vphi^2} \ dx \ . \label{eq:app:NS} \end{eqnarray}
Note that possible dependencies of $F$ and $S$ on $w$ only arise through $\vphi$; thus if $\vphi_w = 0$, necessarily $F_w = 0 = S_w$.

We do now have to compute the $x$ derivatives of $F$ and $S$. This is a matter of standard calculus, and we get
\begin{eqnarray}
\frac{\pa F}{\pa x} &=& - \ \frac{1}{\vphi^2} \ \[ \vphi_t + f \vphi_x - f_x \vphi \ + \ \frac12 \( \s^2 \vphi_{xx} + 2 \s \vphi_{xw} + \vphi_{ww} \) \] \nonumber \\
& & \ - \ \frac{1}{\vphi^2} \ \( \s \s_x \vphi_x + \s_x \vphi_w \)  \ + \ \frac{1}{\vphi^3} \( \s^2 \vphi_x^2 + 2 \s \vphi_x \vphi_w + \vphi_w^2 \) \ , \label{eq:appA:newF} \\
\frac{\pa S}{\pa x} &=& - \ \frac{1}{\vphi^2} \ \( \vphi_w \ + \ \s \, \vphi_x \ - \ \s_x \, \vphi \) \ . \label{eq:appA:newS} \end{eqnarray}
We stress that so far $\vphi$ is a completely arbitrary (twice differentiable)  function.

\subsection{Symmetry as a sufficient condition}
\label{app:symm:suff}

Let us now \emph{assume} that the vector field
$$ Y \ = \ \vphi(x,t;w) \, \pa_x $$ is a (standard) symmetry for the equation \eqref{eq:appA:Ito}. Thus it satisfies the determining equations \eqref{eq:deteq1}, \eqref{eq:deteq2} with $r=0$.

It is immediate to check that in this case it follows immediately from \eqref{eq:appA:newS} that $S_x = 0$.

The situation is slightly more involved for \eqref{eq:appA:newF}; in studying this we have to consider both \eqref{eq:deteq1} and \eqref{eq:deteq2}. It is actually convenient to first consider only \eqref{eq:deteq2}. Using this, we have \begin{eqnarray}
\frac{\pa F}{\pa x} &=& - \ \frac{1}{\vphi^2} \ \[ \vphi_t + f \vphi_x - f_x \vphi \ + \ \frac12 \( \s^2 \vphi_{xx} + 2 \s \vphi_{xw} + \vphi_{ww} \) \] \nonumber \\
& & - \, \frac{\s_x}{\vphi^2} \( \s \vphi_x + \s_x \vphi - \s \vphi_x \) \ + \ \frac{1}{\vphi^3} \( \s^2 \vphi_x^2 + 2 \s \vphi_x (\s_x \vphi - \s \vphi_x ) \right. \nonumber \\
& & \left. + \s_x^2 \vphi^2 - 2 \s \s_x \vphi \vphi_x + \s^2 \vphi_x^2 \) \nonumber \\
&=& - \ \frac{1}{\vphi^2} \ \[ \vphi_t + f \vphi_x - f_x \vphi \ + \ \frac12 \( \s^2 \vphi_{xx} + 2 \s \vphi_{xw} + \vphi_{ww} \) \] \ . \label{eq:app:Fxtransf} \end{eqnarray}
It is then immediate to check that if \eqref{eq:deteq1} holds as well, we have $F_x=0$.

In other words, we have shown that if $\vphi (x,t;w)$ satisfies the determining equations for standard symmetries, see \eqref{eq:deteq1}, \eqref{eq:deteq2}, then the transformed equation is necessarily of the form \eqref{eq:app:target:2}. Moreover, as noted above, if we actually have a deterministic standard symmetry, i.e. $\vphi = \vphi (x,t)$, then $F$ and $S$ do not depend on $w$, and we are in the case of \eqref{eq:app:target}.

\subsection{Symmetry as a necessary condition}
\label{app:symm:nec}

In order to prove that symmetry is not only a sufficient condition but also a necessary one for integrability -- i.e. for the possibility to map the equation into the form \eqref{eq:app:target:2} -- it suffices to run our computations backwards.

In fact, the formulas \eqref{eq:appA:newF} and \eqref{eq:appA:newS} for the $x$ derivatives of the transformed drift and noise coefficients are completely general, and integrability amounts to the existence of a map $\Phi$, i.e. of a generating function -- in the sense of \eqref{eq:app:kozmap} -- $\vphi$ for this.

It is evident that $S_x=0$ implies that $\vphi$ satisfies the second determining equation \eqref{eq:deteq2}. As we have just seen, if this equation is satisfied, then $F_x$ is given by \eqref{eq:app:Fxtransf}, so that the vanishing of $F_x$ is now tantamount to $\vphi$ satisfying the first determining equation \eqref{eq:deteq1} as well.

Finally we note that we can invert \eqref{eq:app:NF} and \eqref{eq:app:NS} to obtain $f$ and $\s$ in terms of $F,S$ and $\vphi$. It is apparent that if $F_w = 0 = S_w$, then it will also be $f_w = 0$, $\s_w = 0$ provided we choose the generating function as $\vphi_w = 0$.

\subsection{Examples}

\medskip\noindent
{\bf Example A.1.} Consider the Ito equation
\beq d x \ = \ \( \frac14 \ + \ \sqrt{x} \) \ dt \ + \ \sqrt{x} \ d w . \eeq
This admits the standard \emph{deterministic} symmetry $Y = \sqrt{x} \pa_x$.
With the change of variable
$$ y \ = \ \int \frac{1}{\vphi} \ dx \ = \ 2 \ \sqrt{x} $$ and using Ito formula, we get that the evolution of $y$ is governed by the Ito equation
\beq d y \ = \ dt \ + \ d w \ . \eeq

\medskip\noindent
{\bf Example A.2.} Consider the Ito equation
\beq d x \ = \ \( x^2 \, + \, x^3 \) \ dt \ + \ x^2 \ d w . \eeq
This admits the standard \emph{deterministic} symmetry $Y = x^2 \pa_x$.
With the change of variable
$$ y \ = \ \int \frac{1}{\vphi} \ dx \ = \ - \, \frac{1}{x} $$ and using Ito formula, we get that the evolution of $y$ is governed by the Ito equation
\beq d y \ = \ dt \ + \ d w \ . \eeq

\medskip\noindent
{\bf Example A.3.} Consider the Ito equation
\beq d x \ = \ e^x \ dt \ + \ d w . \eeq
This admits the standard \emph{random} symmetry $Y = \exp[x-w] \pa_x$.
With the change of variable
$$ y \ = \ \int \frac{1}{\vphi} \ dx \ = \ - \ e^{w-x}  $$ and using Ito formula, we get that the evolution of $y$ is governed by the stochastic differential equation
\beq d y \ = \ e^w \, dt  \ ; \eeq
this is \emph{not} in Ito form.

\medskip\noindent
{\bf Example A.4.} Consider the Ito equation
\beq d x \ = \ \( \frac{x}{2} \ + \ v (t) \) \ dt \ + \ \( x \ + \ u(t) \) \ d w , \eeq
with $u,v$ arbitrary functions.
This admits the standard \emph{random} symmetry $Y = \exp[w] \pa_x$.
With the change of variable
$$ y \ = \ \int \frac{1}{\vphi} \ dx \ = \ - \ e^{-w} \ x   $$ and using Ito formula, we get that the evolution of $y$ is governed by the stochastic differential equation
\beq d y \ = \ e^{-w} \ \[ \(v(t) \ - \ u(t) \)  \, dt \ + \ u(t) \ d w \] \ ; \eeq
this is in general (that is, unless $u(t) = v(t) = 0$) \emph{not} in Ito form.

Choosing e.g. $u(t) = 1 = v(t)$, we get
\beq d y \ = \ e^{- w} \ d w \ . \eeq


\setcounter{equation}{0}

\section{Relations to the Kozlov classification}
\label{app:Kcompare}

As mentioned in the Introduction, a classification of standard deterministic symmetries for scalar Ito equations was also provided by Kozlov \cite{Koz2}; in this Appendix we want to briefly discuss the relations of our classification to Kozlov's one.

We will focus on symmetries for autonomous equations with constant noise,
\beq\label{stoK} d x \ = \ f(x) \, d t \ + \ s \, d w \ . \eeq
We note that Kozlov observes that one can always reduce to this case, and hence discusses a classification of such equations

If we know a symmetry of equation \eqref{stoK}, we can use it to find a change of variable and solve the equation. In particular, the symmetries of equation \eqref{stoK} with different drifts (when there is a symmetry) are given -- by our classification --  in Table \ref{tab:1}.

On the other hand, Kozlov provides a table of results, which we report -- modifying notations so to adapt these to our present ones (note we also separate the two cases with $f(x) = 0$) -- in Table \ref{tab:koz}.

\begin{table}[h]
\begin{center}
\begin{tabular}{||c|c||}
\hline
 $f(x)$ & $\varphi(x,t; w)$\\[3pt]
 \hline
 $0$ & $1$ \\
 \hline
 $0$ & $x$ \\
\hline
$\alpha / x$ &
$x$ \\
\hline
\end{tabular}
\caption{The adapted Kozlov's Table 1 from \cite{Koz2}}
\label{tab:koz}
\end{center}
\end{table}

It is immediate to note that our Table \ref{tab:1} differs in several ways from Kozlov's table, see Table \ref{tab:koz}; it is appropriate to discuss the reason for these differences.

The reasons are three\footnote{Needless to say, there is another difference concerning what is discussed in Remark 9, which concerns W-symmetries; these were completely absent from Kozlov's investigation (again, they were not present in the literature at the time). This difference is not so relevant, as it turns out proper W-symmetries are present only for constant noise (and for the case $\s = s x$, not considered here being already studied in the literature).} (at least)
\begin{enumerate}
\item In Kozlov paper \cite{Koz2}, equations related by a change of variables are in the same class.
\item The Kozlov paper \cite{Koz3} considers $t$ transformations acting on $t$ as well, while we only consider transformations with generator \eqref{eq:Ysimp}.
\item In Kozlov classifications, only \emph{deterministic} standard symmetries are considered, and not \emph{random} ones (for the good reason the latter had not been introduced yet at the time of publication); in other words, the function $\vphi$ being the coefficient of $\pa_x$ in the symmetry vector field is $\vphi = \vphi (x,t)$ in \cite{Koz3}, while in the present work we consider $\vphi = \vphi (x,t;w)$.
\end{enumerate}

The determining equations are, when $r=0$, $\varphi$ does not depend on $w$ and there are no $t$-transformations:
\begin{align}
\varphi_t+f\varphi_x-\varphi f_x+\frac{1}{2}\sigma^2\varphi_{xx}=&0\\
\sigma \varphi_x-\varphi\sigma_x=&0
\end{align}
(remark that in this case, the quotient $\varphi/\sigma$ do not depend on $x$).

Then
\beq y \ = \ \Phi(x,t) \ := \ \int \frac{1}{\varphi(x,t)} \ d x \eeq
and
\beq d y \ = \ \frac{\partial \Phi}{\partial x} \, d x \ + \ \left( \frac{\partial \Phi}{\partial t} \, + \, \frac{1}{2} \, \sigma^2 \, \frac{\partial^2 \Phi}{\partial x^2} \right) \ d t \eeq
that is,
\beq
d y \ = \ \frac{1}{\varphi } \, d x \ + \ \left(- \, \int\frac{\varphi_t}{\varphi^2} \, d x \ - \ \frac{1}{2} \, \sigma^2 \, \frac{\varphi_x}{\varphi^2}\right) \ d t \eeq
Since
\beq d x \ = \ f(x,t) \, d t \ + \ \sigma(x,t) \, d w \eeq
we get (see equation (3.5) in Kozlov \cite{Koz2}):
\beq d y \ = \ \left( \frac{f }{\varphi} \ -\ \int \frac{\varphi_t}{\varphi^2} \, d x \ - \ \frac{1}{2} \, \sigma^2 \, \frac{\varphi_x}{\varphi }\right) \ d t \ + \ \frac{\sigma}{\varphi } \ d w \ .  \eeq

The transformed equation is
\beq d y \ = \ \wt{f}(y,t) \, d t \ + \ \wt{\sigma}(y,t) \, d w \ ,  \eeq
where
\beq
\wt{f}(y,t) \ = \ \frac{f}{\varphi} \, - \, \int\frac{\varphi_t}{\varphi^2} \, d x \ - \ \frac{1}{2} \, \sigma^2 \, \frac{\varphi_x}{\varphi^2}\ , \ \ \  \wt{\sigma}(y,t) \ = \ \frac{\sigma}{\varphi} \ , \eeq
and we can show that these functions do not depend on $y$ ($\frac{\partial x}{\partial y}=\varphi$). In fact,
\begin{eqnarray}
\wt{f}_y &=& -\frac{1}{\varphi}\left(  \varphi_t  +f\varphi_x- f_x \varphi +\frac{1}{2}\sigma^2\varphi_{xx} \right) +\frac{\sigma\varphi_x}{\varphi^2} (\sigma  \varphi_x  - \sigma_x\varphi )=0 \ , \\
\wt{\sigma}_y &=& -\frac{1}{\varphi}( \sigma \varphi_x - \sigma_x \varphi )=0
\ . \end{eqnarray}


\setcounter{equation}{0}

\section{Conservation of symmetries under a change of variables}
\label{app:symmcons}

\def\xt{{\wt{x}}}

As recalled at the beginning of Section \ref{sec:admitted}, standard symmetries are conserved under a change of variables.

It seems this fact was first formally proven by C. Lunini in her Thesis (see \cite{GL1,GL2} for published versions of the result), based on the equivalence between symmetries of an Ito equation and of the corresponding Stratonovich equation, and the transformation properties of Stratonovich equations. We now provide a direct proof of this fact in the scalar case. The computations are similar to those met in \cite{GL1} and in Appendix A to \cite{GSW}, but focused in a different direction.

We note that the result recalled in Remark 3 (quoting \cite{GSW}) mean this preservation of symmetries does not hold for W-symmetries.

\subsection{Change of variables in the Ito equation}

First of all we note that given an Ito equation \eqref{eq:Ito} and a change of variables $\wt{x} = g (x,t)$ with inverse $x = \xi (\wt{x},t)$, the new variable $\wt{x}$ satisfies the Ito equation (also dubbed ``the new equation'')
\beq\label{gen2}
d \wt{x} \ = \ \wt{f} (\wt{x},t ) \, d t \ + \ \wt{\sigma}(\wt{x},t ) \, d w
\eeq
where
\begin{eqnarray}
\wt{f} (\wt{x},t ) &=& \[ \frac{\partial g}{\partial t} \ + \ f \, \frac{\partial g}{\partial x} \ + \ \frac12 \, \sigma^2 \, \frac{\partial^2 g}{\partial x^2} \]_{x = \xi (\wt{x},t)} \ , \label{newf}\\
\wt{\sigma}(\wt{x},t ) &=& \[ \sigma \,  \frac{\partial g}{\partial x} \]_{x = \xi (\wt{x},t)} \ . \label{newsig}
\end{eqnarray}
These equations follow immediately from Ito formula; the necessary computations have been considered in Section \ref{sec:standard} for the special case of $g(x,t)$ as in \eqref{eq:g}.

Note that the change of variables is well defined provided $g_x \not= 0$, and hence $\xi_{wt{x}} \not= 0$; we assume these in the following.

\subsection{Change of variables in the determining equations}

Now we recall that, as seen in Section \ref{sec:firstorder}, the system of determining equations \eqref{eq:deteq1}, \eqref{eq:deteq2} can be rewritten as the equivalent first order system \eqref{eq:R2t}, \eqref{eq:deteq2} (we stress that \eqref{eq:R2t} is equivalent to \eqref{eq:deteq1} \emph{only} on the set of functions satisfying \eqref{eq:deteq2}, which itself depends on $\s$ but not on $f$), where the modified drift $F$ is identified by \eqref{eq:f_frak}.
Repeating here these formulas for ease of reference, we have to deal with the system
\begin{eqnarray}
\vphi_t &+& b \ \vphi_x \ - \ \vphi \ b_x \ + \ r \ \s \ \s_x \ = \ 0 \ , \nonumber \\
\vphi_w &+& \s \ \vphi_x \ - \ \vphi \ \s_x \ - \ r \ \s \ = \ 0 \ ; \label{deteq}\\
& & \ \ b \ := \ f \ - \ \frac12 \ \s \ \s_x \ . \label{frak}  \end{eqnarray}

We now consider the change of variables
$$ \wt{x} \ = \ g(x,t) \ , \ \ \wt{t} \ = \ t \ , \ \ \wt{w}\ = \ w \ , $$
with inverse change of variables
$$ x \ = \ \xi(\wt{x},\wt{t}) \ , \ \ t \ = \ \wt{t} \ , \ \ w=\wt{w} \ . $$
These entail
$$\pa_x \ = \ \( \frac{\pa g}{\pa x} \) \, \pa_{\wt{x}} \ , \ \
\pa_t \ = \ \pa_{\wt{t}} \ + \ \(\frac{\pa g}{\pa t}\) \, \pa_{\wt{x}}  \ , \ \
\pa_w \ = \ \pa_{\wt{w}} \ ;
$$
note that all functions on the r.h.s. should be considered as evaluated in $x = \xi (\wt{x},t)$; in the following we will omit this specification -- even through the notation used in \eqref{newf}, \eqref{newsig} above -- for notational simplicity.

Similarly, we have
$$
\pa_{\wt{x}} \ = \ \( \frac{\pa \xi}{\pa \wt{x}} \) \, \pa_x \ , \ \
\pa_{\wt{t}} \ = \ \pa_t \ + \ \( \frac{\pa \xi}{\pa \wt{t}} \) \, \pa_x \ , \ \
\pa_{\wt{w}} \ = \ \pa_w \ ; $$
here all functions on the r.h.s. should be evaluated in $\wt{x} = g(x,t)$.

Note for later reference that we can write
\beql{gder}
g_x \ = \ \frac{1}{\xi_{\wt{x}}} \ ; \ \  g_t \ = \ - \ \frac{\xi_t}{\xi_{\wt{x}}} \ .
\eeq

This follows at once from considering the Jacobians $J$ and $\wt{J}$ of the direct and inverse changes of coordinates,
$$ J \ = \ \begin{pmatrix} g_x & g_t \\ 0 & 1 \end{pmatrix} \ , \ \ \wt{J} \ = \ \begin{pmatrix} \xi_{\wt{x}} & \xi_t \\ 0 & 1 \end{pmatrix} \ , $$ and recalling that $\wt{J} = J^{-1}$.

Now we pass to consider the vector field
$$ Y \ = \ \vphi (x,t;w) \, \pa_x \ + \ r \, w \, \pa_w \ ; $$
in the new variables this reads
$$ Y \ = \ \[ \vphi (x,t;w) \, \frac{\pa g(x,t)}{\pa x} \]_{x=\xi(\wt{x},t)} \ \pa_{\wt{x}} \ + \ r \, w \, \pa_w \ := \ \wt{\vphi} \, \pa_{\wt{x}} \ + \ r \, w \, \pa_w \ . $$

We have thus defined a new function, $\wt{\varphi}$ given in fully explicit form by
\beq
\wt{\varphi}(\wt{x},t,w) \ := \ g_x [\xi(\wt{x},t),t] \, \varphi[\xi(\wt{x},t),t,w] \ .
\eeq

We do now want to express the symmetry condition for the old equation \eqref{eq:Ito} in terms of the new variables; our task is to show that -- under an additional condition also related to W-symmetries, see below -- this is equivalent to the symmetry condition for the new equation.

We start by expressing the derivatives of $\vphi (x,t;w)$ as derivatives of the function $\wt{\vphi} \( \wt{x} , \wt{t} ; \wt{w} \)$. By standard computations, and using the relations \eqref{gder}, these are
\begin{eqnarray*}
\varphi_{x} &=&\wt{\varphi}_{\wt{x}} \ - \ \( \frac{g_{xx}}{g_x^2} \) \, \wt{\varphi} \nonumber \\
\varphi_t &=& \( \frac{1}{g_x} \)  \, \wt{\varphi}_t \ + \ \( \frac{g_t}{g_x} \) \, \wt{\varphi}_{\wt{x}} \ - \ \( \frac{g_{xt}}{g_x^2} \) \, \wt{\varphi} \label{tphider}\\
\varphi_w &=& \( \frac{1}{g_x} \) \, \wt{\varphi}_w \nonumber
\end{eqnarray*}

Substituting these in the determining equations written in first order form, see \eqref{deteq} above, using the assumption $g_x \not= 0$ (the equations are multiplied by this factor), and recalling it is understood that all functions of $(x,t)$ -- here written within brackets -- should be evaluated in $x = \xi (\wt{x},t)$, we have
\begin{eqnarray}
 \wt{\varphi}_t \ + \ \left( g_x \, b \ + \ g_t \right) \, \wt{\varphi}_{\wt{x}} \ - \ \left(b_x \ + \ \frac{g_{xx}}{g_x} \, b + \frac{g_{xt}}{g_x} \right) \ \wt{\varphi} \ + \ r \, \( g_x \, \s \, \s_x \) &=& 0 \ , \nonumber \\
\wt{\varphi}_w \ + \ \( g_x \, \sigma \) \, \wt{\varphi}_{\wt{x}} \ - \ \left( \frac{g_{xx}}{g_x} \, \sigma \ + \ \sigma_x \right) \ \wt{\varphi} \ - \ r \, \( g_x \, \s \) &=& 0 \ . \nonumber \\
  & &
\end{eqnarray}

Let us define the functions (where $\wt{\sigma}$ should be thought as the new noise)
\begin{eqnarray}
\wt{b
}(\wt{x},t) &=& \[ g_x \, b
 \ + \  g_t \]_{x = \xi (\wt{x},t)} \ , \label{Ftilde} \\
\wt{\sigma}(\wt{x},t) &=& \[ g_x \, \sigma \]_{x = \xi (\wt{x},t)} \ . \label{stilde}  \end{eqnarray}

With this notation, the determining equations \eqref{deteq} for symmetries of the original equation are written in the form
\begin{eqnarray}
\wt{\vphi}_t &+& \wt{b} \ \wt{\vphi}_{\wt{x}} \ - \ \wt{\vphi} \ \wt{b}_{\wt{x}} \ + \ r \ \wt{\s} \ \wt{\s}_{\wt{x}} \ + \ r \, \wt{\s} \[ \wt{\s}_{\wt{x}} \ - \ \( \frac{g_{xx}}{g_x^2} \) \wt{\s} \] \ = \ 0 \ , \nonumber \\
\wt{\vphi}_w &+& \wt{\s} \ \wt{\vphi}_{\wt{x}} \ - \ \wt{\vphi} \ \wt{\s}_{\wt{x}} \ - \ r \ \wt{\s} \ - \ r \, \wt{\s} \ = \ 0 \ ; \label{deteqnew}  \end{eqnarray}
here we have used the identity
$$ \( \wt{\s}_{\wt{x}} \) \ = \ \ \pa_{\wt{x}} \wt{\s} \ = \ \frac{1}{g_x} \pa_x \( g_x \s \) \ = \ \s_x \ + \ \( \frac{g_{xx}}{g_x^2} \) \, g_x \, \s_x \ , $$ which yields
$$ \s_x \ = \ \wt{\s}_{\wt{x}} \ - \ \( \frac{g_{xx}}{g_x^2} \) \ \wt{\s} \ . $$

\bigskip\noindent
{\bf Standard symmetries}

Let us now first look at the case $r=0$. In this case the \eqref{deteqnew} read
\begin{eqnarray}
\wt{\vphi}_t &+& \wt{b} \ \wt{\vphi}_{\wt{x}} \ - \ \wt{\vphi} \ \wt{b}_{\wt{x}} \ + \ r \ \wt{\s} \ \wt{\s}_{\wt{x}} \ = \ 0 \ , \nonumber \\
\wt{\vphi}_w &+& \wt{\s} \ \wt{\vphi}_{\wt{x}} \ - \ \wt{\vphi} \ \wt{\s}_{\wt{x}} \ - \ r \ \wt{\s} \ = \ 0 \ ; \label{deteqnewr0}  \end{eqnarray}
they are the determining equations for symmetries of a transformed equation with noise $\wt{\s} (\wt{x},t)$ and a drift $\wt{f} (\wt{x} , t)$ such that
\beql{newF0} \wt{b} (\wt{x} , t)  \ = \ \wt{f} \ - \ \frac12 \ \wt{\s} \ \wt{\s}_{\wt{x}} \ . \eeq
Our next task is thus to determine what is the $\wt{f}$ corresponding to $\wt{b}$ and $\wt{\s}$ given by \eqref{Ftilde} and \eqref{stilde}.

In fact, recalling how $b$ is written in terms of $f$ and $\s$, the function $\wt{b} (\wt{x} , t)$ is given in terms of $f(x,t)$ and $\sigma (x,t)$ as:
\beql{newF}
\wt{b} \ = \ \[ g_t \ + \ g_x \ \left(f \ - \ \frac12 \, \sigma \, \sigma_x \right) \]_{x = \xi (\wt{x},t)} \ .
\eeq
Hence, requiring \eqref{newF0} to be satisfied means defining $\wt{f}$ as
\beq
\wt{b} \ = \ \wt{f} - \frac12 \, \wt{\sigma} \, \wt{\sigma}_{\wt{x}} \ ,
\eeq
where $\wt{f}$ is new drift, and then
\beq
\wt{f} \ = \ \wt{b} \ + \ \frac12 \, \wt{\sigma} \,\wt{\sigma}_{\wt{x}} \ = \ \( g_t \ + \ g_x \, f \ + \ \frac12 \, \sigma^2 \, g_{xx} \)_{x = \xi(\wt{x},t)}
\eeq
which fits exactly equation \eqref{newf}.

We conclude that: \emph{standard symmetries of an Ito equation are preserved under a change of variables}.

\bigskip\noindent
{\bf Proper W-symmetries}

The discussion conducted above can be repeated in the case with $r\not= 0$; but now the equations \eqref{deteqnew} differ from the determining equations for the transformed equation due to the presence of the extra term
\beq r \ \( \frac{g_{xx}}{g_x^2} \) \ {\wt{\s}}^2 \ = \ r \ g_{xx} \ \s^2 \ . \eeq
As we assumed $r \not= 0$, and of course $\s \not= 0$, this vanishes only for $g_{xx} = 0$. Recalling that $g_x = 1/\s$ and hence $g_{xx} = \s_x/\s^2$, this requirement amounts to $\s_x = 0$, i.e. to requiring that the noise coefficient is constant in space (while it may vary in time).

We conclude that: \emph{proper W-symmetries are preserved under a change of variables if and only if the noise coefficient satisfies $\s_x = 0$}.

This confirms the analysis in \cite{GSclass}.

\setcounter{equation}{0}

\section{Non-existence of W-symmetries for simple non-constant noise}
\label{app:noW}

We have seen in Sect. \ref{sec:const} that in the case of constant noise, $\s (x,t) = s$, there are W-symmetries for cases $A$ and $B$ (no W-symmetry is present for case $C$). We have then stated, in Section \ref{sec:nonconst} (see in particular  Remark 12) that no W-symmetries exist in the case of simple non-constant noise. In this Appendix we prove this assertion.

First of all we note that, as obvious from Remark 9, W-symmetries can exist only if standard symmetries exist. This means we can just consider the cases identified in Section \ref{sec:nonconst}, see in particular Table \ref{tab:4}; obviously we have to bind ourselves to cases $A$ and $B$ in there.

We will consider the $f(x,t)$ identified by our Table \ref{tab:4}, and look for W-symmetries in these cases by considering directly the determining equations \eqref{eq:deteq1} and \eqref{eq:deteq2}.

For a simple noise, $\s (x,t) = s x^k$, the equation \eqref{eq:deteq2} -- which does not depend on $f$, but only on $\s$ -- reads
\beq \vphi_w \ + \ s \, x^k \, \vphi_x \ - \ s \, k \, x^{k-1} \, \vphi \ = \ r \, s \, x^k \ . \eeq
This is a linear (non-homogeneous) equation for $\vphi (x,t;w)$, and is solved by the method of characteristics, yielding (as usual we assume $k\not= 0$, $k \not= 1$)
\beq \vphi (x,t:w) \ = \ x^k \ \psi (z,t) \ - \ \frac{r}{k-1} \ x \ , \eeq
where we have defined
\beq z \ := \ w \ + \ \frac{1}{s \, (k-1)} \ x^{1-k}  \ . \eeq
We could then plug this expression for $\vphi$ into \eqref{eq:deteq1}, obtaining a complex equations which we will not write in general; we will instead consider its form for $f(x,t)$ as provided by the cases $A$ and $B$ of our Table     \ref{tab:4}.

\subsection{Case A}

In case $A$, we have $f(x,t) =  c x^k + (1/2) s^2 k x^{ k - 1}$, and \eqref{eq:deteq1} reads
\beq k \, s^2 \, r \ x^{2 k -1} \ + \ \[ c \, r \ + \ \psi_t \ - \ (c/s) \, \psi_z \] \ x \ = \ 0 \ . \eeq
Having assumed $k \not= 1$, the coefficient of the two different powers of $x$ have to vanish separately, and the vanishing of the coefficient of $x^{2k-1}$ requires, in view of our assumptions $s\not=0$, $k \not= 0$, that $r=0$; this in turn implies there are no proper W-symmetries.

\subsection{Case B}

In case $B$, we have $ f(x,t)  =  c_0 s x^k + (c_1/(1 - k)) x + (1/2) s^2 k x^{2 k - 1}$; the equation \eqref{eq:deteq1} reads then
\beq k \, s^2 \, r \ x^{2 k - 1} \ + \ \frac{c_1}{s \, (k-1)} \ \psi_z \ x \ + \ \[ c_0 \, s \, r \ - \ c_1 \, \psi \ + \ \psi_t \ - \ c_0 \ \psi_z \] \ x^k \ = \ 0 \ . \eeq
As we exclude the case $k=1$, the coefficient of these different powers of $x$ have to vanish separately, and in particular the vanishing f the fist term -- as we assumed $s \not= 0$ and $k \not= 0$ -- requires necessarily $r=0$. Thus no proper W-symmetries are present.

\subsection{A reminder on the multiplicative noise case}

We should stress once again that our result on non-existence of proper W-symmetries for simple noise $\s (x,t) = s x^k$ holds under our assumptions $k \not= 0$ (we have seen in Sect.\ref{sec:const} that in this case there are proper W-symmetries, at least in cases $A$ and $B$) and moreover $k \not= 1$, excluding the case of so called \emph{multiplicative noise}. This assumption is due to the fact such a case has been thoroughly investigated before \cite{GSclass}, studying in detail also the non-autonomous case.

It should be recalled that for multiplicative noise there \emph{are} proper W-symmetries, both in the autonomous and non-autonomous case; see Theorem 1 in \cite{GSclass} -- in particular, cases $(e)$ and $(h)$ therein -- for details.

\setcounter{equation}{0}

\section{Explicit computations in the general case}
\label{app:exagen}

In Sect. \ref{sec:nonconstgen} we stated that the general case -- that is, general noise and non-autonomous equations -- can be dealt with in the way developed here, albeit in that one cannot get an explicit classification.

In this Appendix we want to illustrate the computations needed to discuss such a case; obviously one should focus on a given noise term, and only in this restricted case sufficiently explicit results can be obtained. We will thus consider several examples of this situation, starting from rather trivial ones.

In the simplest example we will give full detail of the computations to make completely clear all of our procedure; in the following examples we will omit trivial details.

\subsection{Example 1}

As a first example, let us consider the case where the noise term is
\beq \s (x,t) \ = \ e^{- t} \eeq (thus the Ito equation becomes asymptotically a deterministic one). In other words we want to determine for which form of the drift term $f(x,t)$ the equation
\beq d x \ = \ f(x,t) \, dt \ + \ e^{- t } \, d w \eeq
admits a symmetry. Note that adding a multiplicative constant to $\s$ and/or a constant in the exponent would lead to equivalent computations.

As per our general procedure, we consider the change of variables
\beq y = \ g (x,t) \ = \ \int \frac{1}{\s(x,t)} \ d x \ , \eeq with inverse $x = \xi (y,t)$. With our choice for $\s$, this means that the direct and the inverse changes of variable are
$$ y \ = \ e^t \ x \ := \ g(x,t) \ ; \ \ \ x \ = \ e^{- t} \, y \ := \ \xi (y,t) \ . $$
By Ito formula, and as usual with $\Delta$ the Ito Laplacian,
\begin{eqnarray*}
dy &=& \frac{\pa g}{\pa x} \, dx \ + \ \frac{\pa g}{\pa t} \, dt \ + \ \frac12 \, \Delta (g) \, dt \\
&=& e^t \, \[ f(x,t) \, dt \ + \ e^{-t} \, dw \] \ + \ e^t \, x \, dt \\
&=& e^t \ \[ f(x,t) \ + \ x \] \ dt \ + \ d w \\
&=& e^t \ \[ f \( \xi (y,t) , t \) \ + \ \xi (y,t) \] \ dt \ + \ d w \\
&=& e^t \ \[ f(e^{-t} y,t) \ + \ e^{-t} y \] \ d t \ + \ d w \\
&:=& F(y,t) \ dt \ + \ d w \ . \end{eqnarray*}

This equation is of the form studied in Sect.\ref{sec:const}, with $s=1$; we know from our discussion there (see also Table 1, where $x$ should be replaced by $y$ and one should set $s=1$) that there can be standard symmetries if and only if $F(y,t)$ is in one of the following forms:
\begin{eqnarray*}
(A) & & F(y,t) \ = \ \Psi_A (y,t) \ = \ H' (t) \ , \\
(B) & & F(y,t) \ = \ \Psi_B (y,t) \ = \ a(t) \ + \ B' (t) \ y \ , \\
(C) & & F(y,t) \ = \ \Psi_C (y,t) \ = \ A' (t) \ + \ b(t) \ \exp [ \b \, y ] \ . \end{eqnarray*}
Correspondingly, the symmetries $Y = \Phi (y,t;w) \, \pa_y$ will be identified by
\begin{eqnarray*}
(A) & & \Phi_A (y,t;w) \ = \ P (y - w - H(t) ) \ , \\
(B) & & \Phi_B (y,t;w) \ = \ \exp [ B(t)] \ , \\
(C) & & \Phi_C (y,t;w) \ = \ \exp[ \b \, (y - w - A(t) )] \ . \end{eqnarray*}
We should also note that
$$ \pa_y \ = \ \( \frac{\pa x}{\pa y} \) \ \pa_x \ . $$
Thus the symmetries will be expressed in the $x$ variable as
\beq Y \ = \ \Phi_a (y,t;w) \, \pa_y \ = \ e^{- t} \ \Phi_a \( g(x,t),t;w\) \ \pa_x \ := \ \vphi_a (x,t;w) \, \pa_x \ . \eeq

We have to express these facts in terms of the original random variable $x$ (rather than $y$). As for $F$, we have seen above that
$$ F(y,t) \ = \ e^t \ \[ f(e^{-t} y,t) \ + \ e^{-t} y \] \ , $$ which can also be written as
$$ F(e^t x , t) \ = \ e^t \ \[ f(x,t) \ + \ x \] \ ; $$ this in turn implies
\beq f(x,t) \ = \ e^{-t} \ F(e^t x , t) \ - \ x \ . \eeq
We have then to apply this formula to the three admissible cases for $F(y,t) = \Psi_a (y,t)$ enumerated above. We obtain
\begin{eqnarray*}
(A) & & f(x,t) \ = \ f_A (x,t) \ = \ e^{-t} \, H' (t) \ - \ x \ , \\
(B) & & f(x,t) \ = \ f_B (x,t) \ = \ e^{-t} \, a(t) \ + \ x \, B' (t) \ - \ x \ , \\
(C) & & f(x,t) \ = \ f_C (x,t) \ = \ e^{-t} \ \[ A' (t) \ + \ \exp[ \b \, e^t \, x ] \ b(t)\] \ - \ x \ . \end{eqnarray*}

Correspondingly, the coefficients of the symmetry vector fields expressed in the $x$ variable are
\begin{eqnarray*}
(A) & & \vphi_A (x,t;w) \ = \ e^{-t} \ P [ e^t x - w - H (t) ] \ , \\
(B) & & \vphi_B (x,t;w) \ = \ \exp[ - t \ + \ B(t) ] \ , \\
(C) & & \vphi_C (x,t;w) \ = \ \exp [ - t \ + \ \b \, ( e^t x - w - A(t) ) ] \ . \end{eqnarray*}

One can check by direct computations that these $\vphi_a$, together with the $F_a$ given above (where $a=A,B,C$), satisfy the determining equations \eqref{eq:deteq1}, \eqref{eq:deteq2}.

\subsection{Example 2}

As a second Example, we consider the case of an autonomous equation, with a non-simple noise. More specifically, we consider
\beq \s (x) \ = \ \sin^2 (x) \ . \eeq
In this case, we want to restrict our attention to \emph{autonomous} Ito equation; that is, the drift coefficient will also be assumed to be a function of $x$ alone, $f = f(x)$, and we consider equations of the form
\beq dx \ = \ f(x) \, dt \ + \ \sin^2 (x) \, d w \ . \eeq
In this case we have
$$ y \ = \ g(x) \ = \ \int \frac{1}{\s (x)} \ dx \ = \ - \, \cot (x) \ ; $$
the inverse change of coordinates is
$$ x \ = \ \xi (y) \ = \ - \, \mathrm{arccot} (y) \ ; \ \ \frac{d \xi}{d y} \ = \  \frac{1}{1 + y^2} \ . $$
Proceeding as above, we get
\begin{eqnarray*}
dy &=& \frac{\pa g}{\pa x} \, dx \ + \ \frac{\pa g}{\pa t} \, dt \ + \ \frac12 \, \Delta (g) \, dt \\
&=& \( \csc^2(x) f(x)-\cos (x) \sin (x) \) \ dt \ + \ d w \\
&:=& F(y) \, dt \ + \ d w \ . \end{eqnarray*}
In the last step it is understood that $x$ is seen as $x = \xi (y)$.

Thus $f(x)$ is expressed in terms of $F(x)$ as
\beq f(x) \ = \ \sin^2 (x) \ \[ F[ g(x)] \ + \ \sin (x) \, \cos (x) \] \ . \eeq
We get, with the same notation as above,
\begin{eqnarray*}
f_A (x) &=& \sin^2 (x) \ \[ c \ + \ \sin (x) \, \cos (x) \] \ , \\
f_B (x) &=& \sin^2 (x) \ \[ c_0 \ - \ c_1 \, \cot (x) \ + \ \sin (x) \, \cos (x) \] \ , \\
f_C (x) &=& \sin^2 (x) \ \[ c_0 \ + \ c_1 \, \exp [ - \b \, \cot (x) ] \ + \ \sin (x) \, \cos (x) \] \ . \end{eqnarray*}

Similarly, proceeding as in Example 1, we obtain that
\begin{eqnarray*}
\vphi_A (x,t;w) &=& \sin^2 (x) \ P \[ - \cot (x) - w - ct \] \ , \\
\vphi_B (x,t;w) &=& \sin^2 (x) \ \exp [ c_1 \ t ] \ , \\
\vphi_C (x,t;w) &=& \sin^2 (x) \ \exp[ - \b \, ( \cot (x) + w + c_0 t ) ] \ . \end{eqnarray*}

Direct substitution in \eqref{eq:deteq1}, \eqref{eq:deteq2} shows again that indeed $f_a$ and $\vphi_a$ satisfy the determining equations for the $\s(x)$ we are considering.

\subsection{Example 3}

Finally, let us consider a case where $\s$ does actually depend on both $x$ and $t$. We choose
\beq \s (x,t) \ = \ e^{-t} \ \sqrt{x} \ . \eeq
In this case we get
$$ y \ = \ g(x,t) \ = \ 2 \ \sqrt{x} \ e^t \ , \ \ \xi (y,t) \ = \ \frac14 y^2 \ e^{- 2 t} \ , \ \ \frac{\pa \xi (y,t)}{\pa y} \ = \ e^{- t} \ \sqrt{x} \ . $$

Using Ito formula as above, we get
\begin{eqnarray*}
dy &=& \( \frac{e^{-t} \ \left(8 \, e^{2 t} \, x \ + \ 4 \, e^{2 t} \,
   f(x,t) \ - \ 1 \right)}{4 \ \sqrt{x}} \) \ dt \ + \ d w \\
   &:=& F(y,t) \ dt \ + \ d w \ . \end{eqnarray*}
This yields
\beq f(x,t) \ = \ \frac14 \ e^{- 2 t} \ \[ 1 \ + \ 4 \, e^t \, F[g(x,t),t] \, \sqrt{x} \ - \ 8 \, e^{2 t} \, x \] \ . \eeq
The usual procedure gives
\begin{eqnarray*}
f_A (x,t) &=& \frac14 \ e^{- 2 t} \ \( 1 \ + \ 4 \, e^t \, \sqrt{x} \, H' (t) \ - \ 8 \, e^{2 t} \, x \) \ , \\
f_B (x,t) &=& \frac14 \ e^{- 2 t} \ \( 1 \ + \ 4 \, e^t \, \sqrt{x} \( a(t) \ + \ 2 \, e^t \, \sqrt{x} \, B' (t) \) \ - \ 8 \, e^{2 t} \, x \) \ , \\
f_C (x,t) &=& \frac14 \ e^{- 2 t} \ \(1 \ + \ 4 \, e^t \, \sqrt{x} \, \( A' (t) \ + \ b(t) \, \exp[2 \, \b \, e^t \, \sqrt{x} ] \) \ - \ 8 \, e^{2 t} \, x \) \ ; \\
\vphi_A (x,t;w) &=& e^{-t} \, \sqrt{x} \ P[ 2 \, e^t \, \sqrt{x} \ - \ w \ - \ H(t) ] \ , \\
\vphi_B (x,t;w) &=& e^{-t} \, \sqrt{x} \ e^{B(t)}  \ , \\
\vphi_C (x,t;w) &=& e^{-t} \, \sqrt{x} \ \exp \[- \, \b \, \( A(t) \ + \ w \ - \ 2 \, e^t \, \sqrt{x} \) \]  \ . \end{eqnarray*}

Once again, direct substitution in \eqref{eq:deteq1}, \eqref{eq:deteq2} shows that $\s (x,t)$, $f_a (x,t)$ and $\vphi_a (x,t;w)$ satisfy the determining equations.

\subsection{Example 4}

In our previous Examples we have actually performed the computations taking the Ito equation to its standard form, imposed them to be of the symmetric type, and verified that these lead to the expected form for the original drift coefficients. In other words we have verified that our general construction leads, when applied to equations with the considered noise terms, to the equations predicted in general by Proposition 3 and in the autonomous case by Proposition 4.

The real use of the Propositions given in Section \ref{sec:nonconstgen}, however, is just to \emph{avoid} these computations, providing a closed form formula for their result.

Thus we consider e.g. the noise term
\beq \s (x,t) \ = \ \frac{\sqrt{x}}{1 \, + \, t^2} \ . \eeq
Then Proposition 4 states that the drift terms -- and the corresponding symmetry vector field coefficients -- are given by
\begin{eqnarray*}
f_A (x,t) &=& \frac{1}{4 (t^2+1)^2} \ \( 1 -16 x t^3+4 \sqrt{x} H'(t) t^2-16
   x t+4 \sqrt{x} H'(t) \) \ , \\
f_B (x,t) &=& \frac{1}{4 (t^2+1)^2} \ \[ 1 -16 x t^3+4
   \sqrt{x} \left(a(t)+2 \left(t^2+1\right)
   \sqrt{x} B'(t)\right) t^2 \right. \\
   & & \left. -16 x t+4 \sqrt{x}
   \left(a(t)+2 \left(t^2+1\right) \sqrt{x}
   B'(t)\right) \] \ , \\
f_C (x,t) &=& \frac{1}{4 (t^2+1)^2} \ \[  1 -16 x t^3+4
   \sqrt{x} \left(e^{2 \beta
   \left(t^2+1\right) \sqrt{x}}
   b(t)+A'(t)\right) t^2 \right. \\
   & & \left. -16 x t+4 \sqrt{x}
   \left(e^{2 \beta  \left(t^2+1\right)
   \sqrt{x}} b(t)+A'(t)\right) \] \ ; \\
\vphi_A (x,t;w) &=& \frac{\sqrt{x} \ P\left[2 \sqrt{x}
   \left(t^2+1\right)-w-H(t)\right]}{t^2+1} \ , \\
\vphi_B (x,t;w) &=& \frac{e^{B(t)} \sqrt{x}}{t^2+1} \ , \\
\vphi_C (x,t;w) &=& \frac{e^{\beta
   \left(2 \sqrt{x}
   \left(t^2+1\right)-w-A(t)\right)}
   \sqrt{x}}{t^2+1} \ . \end{eqnarray*}
As usual, checking that the $\vphi_a (x,t;w) $ are indeed symmetries of the Ito equations with drift $f_a (x,t)$ and the considered noise term amounts to a standard computation.

 It goes without saying that a direct classification of the Ito equations with this noise term admitting symmetries, and of the corresponding symmetries, would require a substantial effort.


\label{lastpage}
\end{document}